\documentclass{aa}  

\usepackage{placeins}
\usepackage{graphicx}
\usepackage{txfonts}

\def\PLUTO{{\sc pluto}}

\newcommand{\msun}{M$_{\odot}$}
\newcommand{\mej}{$M_{\rm ej}$}
\newcommand{\ebw}{$E_{\rm bw}$}
\newcommand{\tcrb}{T\,CrB}

\begin{document}

   \title{Predicting the X-ray signatures of the imminent T~Coronae Borealis outburst through 3D hydrodynamic modeling}

   \author{S. Orlando\inst{1}
         \and L.\,Chomiuk\inst{2}
         \and J.\,J.\,Drake\inst{3}
         \and M.\,Miceli\inst{4,1}
         \and F.\,Bocchino\inst{1}
         \and O.\,Petruk\inst{1,5}
         }

         \institute{INAF -- Osservatorio Astronomico di Palermo, Piazza del Parlamento 1, I-90134 Palermo, Italy\\
         \email{salvatore.orlando@inaf.it}
         \and Center for Data Intensive and Time Domain Astronomy, Department of Physics and Astronomy, Michigan State University, East Lansing, MI 48824, USA
         \and Lockheed Martin Solar and Astrophysics Laboratory, 3251 Hanover St, Palo Alto, CA 94304, USA
         \and Dip. di Fisica e Chimica E. Segr\`e, Universit\`a degli Studi di Palermo, Piazza del Parlamento 1, 90134 Palermo, Italy
         \and Institute for Applied Problems in Mechanics and Mathematics, Naukova Str. 3-b, 79060 Lviv, Ukraine
         }

   \date{Received September 15, 1996; accepted March 16, 1997}

  \abstract
   {T~Coronae~Borealis (\tcrb) is a symbiotic recurrent nova, with confirmed eruptions in 1866 and 1946. Mounting evidence suggests an imminent outburst, offering a rare opportunity to observe a nearby nova.}
   {We constrain the circumbinary medium (CBM) properties by modeling inter-eruption radio data, then simulate the hydrodynamic evolution of the upcoming \tcrb\ outburst to predict its X-ray signatures, focusing on the impact of the red giant companion, accretion disk, and equatorial density enhancement (EDE) on the remnant and emission.}
   {We model the thermal radio signatures of a CBM consisting of a spherical wind-like component and a torus-like equatorial density enhancement to quantify its density. We then run three-dimensional hydrodynamic simulations of the nova outburst, varying explosion energies, ejecta masses, and circumbinary configurations. From these, we synthesize X-ray lightcurves, spectra, and maps as observed by XMM-Newton and XRISM.}
   {The CBM in \tcrb\ is much less dense than in other symbiotic recurrent novae, with a mass-loss rate of $\dot{M}\approx4\times 10^{-9}$~M$_{\odot}$~yr$^{-1}$ for a 10~km~s$^{-1}$ wind. Despite the expected low-density CBM, the outburst's blast will be strongly collimated along the poles by the combined influence of the accretion disk and EDE, producing a bipolar shock. The companion star partially shields the ejecta, forming a bow shock and a hot wake. The X-ray evolution proceeds through three phases: an early phase (first few hours) dominated by shocked disk material; an intermediate phase ($\sim 1$ week to 1 month) driven by reverse-shocked ejecta; and a late phase dominated by shocked EDE. Soft X-rays trace shocked ejecta, while hard X-rays originate from shocked ambient gas. Synthetic spectra show asymmetric, blueshifted lines due to absorption of redshifted emission by expanding ejecta.}
   {The predicted X-ray evolution of \tcrb\ shows similarities to RS\,Oph and V745\,Sco, reaching a comparable peak luminosity ($L_\mathrm{X}\approx 10^{36}\,\mathrm{erg\,s^{-1}}$), but with a more prolonged soft X-ray phase, reflecting its less dense CBM and distinct ejecta–environment interaction.}


\titlerunning{Predicting the X-ray signatures of the imminent T~Coronae Borealis outburst}
\authorrunning{S. Orlando et~al.}

   \maketitle
%

\section{Introduction}

T Coronae Borealis (\tcrb; also known as HD~143454 or HR~5958) is a prototypical symbiotic recurrent nova system located at a distance of $\approx 890$\,pc (e.g., \citealt{2021AJ....161..147B}), consisting of a massive white dwarf (WD) accreting hydrogen-rich material from a red giant (RG) companion. Historically, \tcrb\ has undergone two confirmed thermonuclear outbursts, in 1866 and 1946 (with additional possible eruptions recorded in 1217 and 1787; \citealt{2023JHA....54..436S}), both exhibiting the characteristic signatures of a nova explosion triggered by a thermonuclear runaway on the surface of the WD (e.g., \citealt{2000NewAR..44...81S}). These include a rapid rise in optical brightness, followed by a complex evolution involving shocks, radiation-driven winds, and a prolonged decline phase.

Nearly eight decades have passed since the last eruption, and \tcrb\ has remained in a state of quiescence, albeit with increasingly active behavior across the electromagnetic spectrum. Notably, the interval between the 1866 and 1946 outbursts was 80 years, suggesting that, if a similar recurrence timescale applies, the next eruption could occur around 2026. This timing aligns with a growing body of observational and theoretical evidence accumulated over the past decade, indicating that \tcrb\ is approaching the critical conditions for another imminent outburst (e.g., \citealt{2016NewA...47....7M, 2023AstL...49..501M, 2023RNAAS...7..251M, 2023RNAAS...7..145M, 2023JHA....54..436S}). This includes long-term increases in accretion rate, quasi-periodic optical brightenings, and enhanced high-energy emission (e.g., \citealt{2018A&A...619A..61L, 2025A&A...694A..85P}), all interpreted as possible precursors of an impending nova event. If confirmed, the upcoming eruption of \tcrb\ would offer an unprecedented opportunity to witness, in real time, the onset and development of a thermonuclear nova in a well-studied, nearby symbiotic system.

Despite extensive historical data and ongoing monitoring, key questions remain unresolved, including the structure and density of the circumbinary medium (CBM), the geometry and dynamics of the ejecta, and the role of binary interactions in shaping the nova remnant. Recent theoretical studies have offered new insights into the evolution of \tcrb. \citet{2025ApJ...982...89S} performed one-dimensional (1D) hydrodynamic (HD) simulations exploring mass-accretion rates onto $\sim1.35\,M_\odot$ WD with carbon-oxygen or oxygen-neon cores. Their models reproduce ignition after $\sim$80 years of accretion and suggest that the WD gains mass after each thermonuclear runaway, potentially reaching the Chandrasekhar limit and leading to either a Type~Ia supernova (SN) or accretion-induced collapse. Complementary work by \citet{2025A&A...698A.251J} examined how variations in WD mass, initial luminosity, metallicity, and accretion rate affect outburst properties. They find that recurrence on $\sim$80-year timescales requires accretion rates of $\dot{M}_{\mathrm{acc}}\sim10^{-8}$--$10^{-7}\,M_\odot\,\mathrm{yr}^{-1}$ onto massive WDs ($M_{\mathrm{WD}}\sim1.30$--$1.38\,M_\odot$) with luminosities $L_{\mathrm{WD}}\sim0.01$--$1\,L_\odot$. Lower metallicity or luminosity leads to higher accumulated masses and ignition pressures, resulting in more energetic eruptions.

Together, these theoretical efforts provide a solid framework for interpreting the forthcoming outburst of \tcrb, offering predictions for its energetics, recurrence timescale, and chemical composition of the ejecta. However, they do not capture the inherently multidimensional and asymmetric nature of nova explosions and their interaction with the surrounding medium, highlighting the need for full three-dimensional (3D) models.

Recent advances in computational astrophysics have made it possible to simulate nova explosions in full 3D, taking into account the interaction with anisotropic CBM structures, such as the equatorial density enhancement (EDE) commonly inferred in symbiotic systems. Analogous studies of other recurrent novae, including RS\,Oph, U\,Sco, V745\,Sco, and V407\,Cyg, have demonstrated that the presence of an EDE can significantly affect the morphology of the remnant, leading to bipolar shock structures, asymmetric X-ray emission, and complex absorption profiles (e.g., \citealt{2008A&A...484L...9W, 2009A&A...493.1049O, 2010ApJ...720L.195D, 2012MNRAS.419.2329O, 2015ApJ...806...27P, 2017MNRAS.464.5003O}). These simulations have successfully reproduced the asymmetric emission line profiles seen in high-resolution spectra of the early phases of nova blast waves \citep[e.g.][]{2009ApJ...691..418D,Drake2016}.

In this study, we present the first 3D HD simulations aimed at predicting the morphological and observational signatures of the anticipated \tcrb\ outburst. Our models incorporate realistic parameters derived from recent observations (see Table~\ref{tab1}), including the binary separation (\citealt{1998MNRAS.296...77B}), the size of the RG companion (\citealt{2025ApJ...983...76H}), the density and spatial profile of the accretion environment, and plausible ranges of explosion energy and ejected mass. While some minor discrepancies exist among published parameters (e.g., \citealt{2025A&A...694A..85P}), they are sufficiently small to have negligible impact on our results. We consider both scenarios with and without an accretion disk, and we explore the impact of varying the EDE's thickness and density on the resulting blast wave evolution.

A primary objective of this work is to identify the key parameters that govern the degree of collimation of the blast and the resulting asymmetries in the X-ray emission. By synthesizing emission maps and spectra as they would appear to current instruments (e.g., XMM-Newton and XRISM), we aim to provide a predictive framework for interpreting forthcoming observations. Special attention is devoted to the expected distribution of emission measure (EM) versus temperature and ionization time, the development of forward and reverse shocks, and the imprint of the disk, the EDE, and the RG companion on the shock morphology.

Given the likely imminence of \tcrb’s eruption, the models presented here are intended to guide the community’s observing strategies across multiple wavelengths and platforms. The predictions derived from this study are not only useful for understanding the physical processes in \tcrb\ itself but also have broader implications for the population of recurrent novae and their potential role as progenitors of Type Ia SNe.

This paper is organized as follows: Sect.~\ref{sec:cbm} describes the model of the thermal radio signatures of a CBM; Sect.~\ref{sec:model} outlines the physical assumptions of our HD models for the nova eruption; Sect.~\ref{sec:results} presents the blast wave evolution, remnant morphology, and X-ray emission; Sect.~\ref{sec:discussion} discusses constraints on ejected mass and explosion energy, and compares results with RS~Oph and V745~Sco; Sect.~\ref{sec:summary} summarizes our conclusions. Appendix~\ref{app:setup} describes the computational setup; Appendix~\ref{app:synthesis} details X-ray synthesis; Appendices~\ref{sec:em_dist} and~\ref{app:add_spec} provide additional results, including EM distributions and spectra.

\section{Modeling radio constraints on the CBM}
\label{sec:cbm}

Few published studies exist on the properties of the CBM in \tcrb, despite it being critical for determining shock signatures during eruption. However, constraints can be inferred from radio observations of the system during its "super-active" inter-eruption state in 2016 (\citealt{2019ApJ...884....8L}). The radio emission from symbiotic stars is generally attributed to free-free thermal emission from the ionized CBM, which is photoionized by the accreting WD (\citealt{1984ApJ...284..202S, 1984ApJ...286..263T}). During the super-active state, the CBM in \tcrb\ is thought to be largely ionized (\citealt{2016NewA...47....7M}), meaning that the observed radio emission should offer a reasonably complete measure of the  mass and the EM of the CBM. To quantify the CBM density, we consider the brightest epoch of the Karl G.\ Jansky Very Large Array (VLA) monitoring observations of \cite{2019ApJ...884....8L}, which occurred on 2016 July 14. At 5 GHz, the flux density was measured to be $0.075\pm0.014$~mJy, and at 35 GHz, $0.513\pm0.038$~mJy. In addition, on 2016 Dec 19, \tcrb\ was observed in the VLA's high resolution A configuration, when it was observed to be an unresolved point source, with a resolution of 0.07 arcsec. In modeling these observations, we take a distance to \tcrb\ of $D = 890$~pc (\citealt{2021AJ....161..147B}).

To constrain the CBM with these radio observations, the simplest model is a  spherically symmetric, wind-like CBM, with a density profile resulting from mass loss at a rate $\dot{M}$ and terminal wind velocity $v_{\infty}$ of:

\begin{equation} \label{eq:denswind}
n_{\mathrm{w}} = \frac{\dot{M}}{4\pi r^2 \mu m_{\mathrm{H}} v_{\infty}}\,.
\end{equation}

\noindent
Eq.~2 of \citet{1990ApJ...349..313S} links the measured radio flux density to the wind density, and we assume $v_{\infty} = 10$~km~s$^{-1}$. The 5~GHz flux density measured by \cite{2019ApJ...884....8L} implies a relatively low wind density, corresponding to a mass-loss rate of $\dot{M} \approx 4 \times 10^{-9}$~M$_{\odot}$~yr$^{-1}$. This value reflects the portion of the RG wind that remains in the circumbinary environment and contributes to the radio emission, rather than the total mass lost by the RG, as some of the wind material is likely accreted onto the WD. In the HD models of \tcrb's eruption (Sect.~\ref{sec:model}), we  explored wind densities, $n_{\rm w}$ at 1 pc, ranging from 1 to $5 \times 10^{-3}$~cm$^{-3}$ (see Table~\ref{tab2}). These values correspond to mass-loss rates in the range between $4\times 10^{-9}$ and $2\times 10^{-8}$~\msun~yr$^{-1}$ (we consider higher density winds here, motivated by the possibility that the CBM was not entirely ionized in 2016). Notably, the upper end of this range aligns with alternative estimates provided in the literature (e.g., \citealt{2023A&A...680L..18Z, 2024MNRAS.532.1421T}). 

A $\dot{M} = 4 \times 10^{-9}$~M$_{\odot}$~yr$^{-1}$ places \tcrb\ among the symbiotic stars with the lowest radio luminosities and mass-loss rates in the \cite{1990ApJ...349..313S} sample, highlighting the relatively sparse CBM in this system despite its periodic nova outbursts.  
How does the CBM of \tcrb\ compare to those of other symbiotic recurrent novae? To address this question, we contrast \tcrb\ with RS~Oph and V745~Sco, two other well-studied symbiotic recurrent novae. For RS Oph, we examine an inter-eruption quiescent observation obtained with the VLA by R.\ Hjellming (NRAO program code AH573) on 1996 May 29, 11 years after the 1985 eruption.  RS~Oph was detected with flux density of $0.41\pm0.08$ mJy at 8.4 GHz. At an assumed distance of 2.7 kpc (\citealt{2022MNRAS.517.6150S})  and assuming a $S_{\nu} \propto \nu^{0.6}$ spectral index \citep{Panagia75,Wright75}, this corresponds to a mass-loss rate of $\dot{M} \approx 6 \times 10^{-8}$ M$_{\odot}$ yr$^{-1}$, more than an order of magnitude higher than the value inferred for \tcrb. For V745~Sco, although no radio detections exist during quiescence (\citealt{2024MNRAS.534.1227M}), modeling of its nova eruption suggests a pre-outburst mass-loss rate of $\dot{M} \approx 2 \times 10^{-8}$ M$_{\odot}$ yr$^{-1}$ (\citealt{2017MNRAS.464.5003O, 2024MNRAS.534.1227M}), again substantially denser than the CBM inferred for \tcrb. These comparisons reinforce the conclusion that \tcrb\ has an unusually low-density CBM among symbiotic recurrent novae. 

However, the circumbinary environment is likely more complex than a simple spherical wind. HD simulations suggest that some of the RG wind may be gravitationally focused into an EDE by interaction with the WD companion (e.g., \citealt{2007ASPC..372..397M, 2016MNRAS.457..822B}). If this EDE is ionized, it would also contribute to the thermal radio flux during quiescence, particularly since the characteristic size of the EDE may exceed that of the radio photosphere of a spherical stellar wind, for parameters we are considering in \tcrb\ (Table~\ref{tab2}). We modeled the CBM with a combined spherical wind and EDE, with the density described as (e.g., \citealt{2017MNRAS.464.5003O}):

\begin{equation}
\rho(\mathbf{r}) = \rho_{\mathrm{w}} \left( \frac{1~\mathrm{pc}}{r} \right)^2 + \rho_{\mathrm{ede}} \exp\left[-\left( \frac{x}{h_x} \right)^2 - \left( \frac{y}{h_y} \right)^2 - \left( \frac{z}{h_z} \right)^2 \right]
\label{eq2}
\end{equation}

\noindent
where \( \rho(\mathbf{r}) \) is the mass density at position \( \mathbf{r} = (x, y, z) \), \( r \) is the radial distance from the center of the RG, \( \rho_{\mathrm{w}} = \mu m_{\mathrm{H}} n_{\mathrm{w}} \) is the wind density at a reference distance of 1 pc, \( \rho_{\mathrm{ede}} = \mu m_{\mathrm{H}} n_{\mathrm{ede}} \) is the central density of the EDE, \( \mu \approx 1.3 \) is the mean atomic weight, \( m_{\mathrm{H}} \) is the mass of a hydrogen atom, \( n_{\mathrm{w}} \) is the hydrogen number density of the stellar wind at 1 pc, \( n_{\mathrm{ede}} \) is the peak hydrogen number density of the EDE, \( h_x, h_y, h_z \) are the characteristic scale lengths of the EDE in the \( x, y, z \) directions, respectively. We also took a binary inclination (and inclination of the EDE) of $55^{\circ}$ (\citealt{2025ApJ...983...76H}; see also Table~\ref{tab1}). 

To model the thermal free-free radio emission from this CBM, we implemented this density distribution as a 3D array, rotated the cube to an inclination of $55^{\circ}$, and integrated the cube along the line of sight, to infer the thermal optical depth at frequency $\nu$ \citep{Lang80}:  

\begin{equation}
\tau_{\nu} = 8.235 \times 10^{-2} \left(\frac{T_{\rm e}}{\rm K}\right)^{-1.35} \left(\frac{\nu}{\rm GHz}\right)^{-2.1} \left(\frac{{\rm EM}_{\rm l}}{\rm cm^{-6}\,pc}\right)  
\end{equation}

\noindent
where $T_{\rm e}$ is the electron temperature (assumed to be $10^4$~K in our models, consistent with photoionization from the WD) and EM$_{\rm l}$ is the path-length emission measure (relevant in the radio, in contrast with the volume emission measure that determines X-ray emission measure, EM). The resulting radio flux density $S_{\nu}$ from each ``pixel" of the simulation is:

\begin{equation}
S_{\nu} = \frac{2 k T_{\rm e} \nu^2}{c^2} \left(\frac{l}{D}\right)^2 \left(1 - e^{-\tau_{\nu}}\right)
\end{equation}

\noindent
\citep{Seaquist&Bode08}  in cgs units, where $k$ is the Boltzmann constant, $c$ is the speed of light, and $l$ is the width of one of the simulation cells. 

We modeled the radio emission from the same CBM distributions considered in our HD models (Table~\ref{tab2}), assuming the entire CBM is singly ionized across all elements. We find that models with $n_{\rm w} =  10^{-3}$~cm$^{-3}$ and $n_{\rm ede} = 10^6$~cm$^{-3}$ are consistent with observed radio constraints. For example, our fiducial model (Run 4: M3-E3-D6-Z4.5-X4-W1) would yield a 5 GHz flux density of 0.12 mJy (slightly higher than observed on 2016 July 14) and a 35 GHz flux density of 0.28 mJy (slightly lower than observed). This model of the CBM is also consistent with radio constraints on the apparent size of \tcrb\ (predicted to be 3 mas in diameter at 35 GHz). 
The expected radio flux densities mildly increase for a larger EDE scale length (Run 8: M3-E3-D6-Z3-X8-W1), to 0.21 mJy at 5 GHz and 0.35 mJy at 35 GHz, still consistent with observations.
Models of the EDE with $n_{\rm ede} = 10^7$~cm$^{-3}$ predict significantly higher flux densities, inconsistent with radio constraints if the full EDE is ionized. For example, Runs 5 and 10--13 predict a 5 GHz flux density of 2.2 mJy and a 35 GHz flux density of 11 mJy, significantly higher than observed by \cite{2019ApJ...884....8L}. Models with $n_{\mathrm{ede}} = 10^8$~cm$^{-3}$ are even more luminous (4.6 mJy at 5 GHz and a whopping 123 mJy at 35 GHz for Run 9, M3-E3-D8-Z3-X8-W1), and would also have been easily resolvable with the VLA at 35 GHz ($0.09 \times 0.16$ arcsec in diameter).

Based on this analysis, we prefer for \tcrb\ models of the CBM with low densities, $n_{\rm w} =  10^{-3}$~cm$^{-3}$ and $n_{\rm ede} = 10^6$~cm$^{-3}$ (Runs 2, 4, 8, and 14). Nevertheless, our HD modeling also explores scenarios with a denser CBM, motivated by the possibility that the medium was not fully ionized, even during the super-active accretion phase of 2016. This approach is further supported by alternative density estimates reported in the literature (e.g., \citealt{2023A&A...680L..18Z, 2024MNRAS.532.1421T}) and by evidence from ALMA measurements indicating that the wind was far from fully ionized in 2024 (\citealt{Petry2025}).

\section{3D hydrodynamic modeling of eruption}
\label{sec:model}

To investigate the expected outburst of \tcrb, we have developed a 3D  HD model that captures the expansion of the blast wave as it propagates through the complex circumbinary environment of this symbiotic binary system. Our simulation framework builds on previous efforts to model analogous systems (most notably RS\,Oph and V745\,Sco) where collimated blast morphologies and asymmetric emission features have been successfully interpreted using 3D simulations (e.g., \citealt{2009A&A...493.1049O, 2017MNRAS.464.5003O}). In the following, we describe the adopted physical model and the exploration of key parameters that influence the outburst evolution. Appendix~\ref{app:setup} provides further details on the numerical setup and the code used, while Appendix~\ref{app:synthesis} explains the methodology employed to synthesize the X-ray emission from the evolving nova remnant.

In our model, we describe a binary system that comprises a WD with a mass of $1.37\,M_{\odot}$ (e.g., \citealt{2004A&A...415..609S, 2025ApJ...983...76H}), accreting material from a RG companion. In the literature, estimates of the RG's mass range from $M_{\rm rg} = 0.69\,M_{\odot}$ (\citealt{1998MNRAS.296...77B, 2025ApJ...983...76H}) to $M_{\rm rg} = 1.12\,M_{\odot}$ (\citealt{2004A&A...415..609S}), while its radius is consistently estimated at approximately\footnote{\cite{2025A&A...694A..85P} reports a slightly larger radius for the RG companion ($\approx 73 \pm 8.6$\,\msun). However, the value adopted in this study lies within $1 \sigma$ of that estimate.} $R_{\rm rg} = 64\,R_{\odot}$ (e.g., \citealt{2025ApJ...983...76H}). Since our simulations do not include gravitational forces, the specific values of $M_{\rm wd}$ and $M_{\rm rg}$ do not affect the simulation outcomes. Nevertheless, for consistency with recent observational constraints, we adopt $M_{\rm rg} = 0.69\,M_{\odot}$ and $R_{\rm rg} = 64\,R_{\odot}$, following \cite{2025ApJ...983...76H}. As for the orbital separation, there is broad agreement in the literature that it is approximately 0.9\,au (e.g., \citealt{1998MNRAS.296...77B}), though some studies suggest slightly smaller values, around 0.5\,au (\citealt{2000AJ....119.1375F, 2019ApJ...884....8L}). The full set of parameters adopted in our simulations is summarized in Table~\ref{tab1}. 

\begin{table}
\caption{Input parameters for the binary system model.}
\label{tab1}
\begin{center}
\begin{tabular}{llll}
\hline
\hline
Parameter    & Description         & Value             & Reference \\ \hline
$M_{\rm wd}$ & WD mass    & $1.37\,M_{\odot}$ &  a, b\\
$M_{\rm rg}$ & RG mass      & $0.69\,M_{\odot}$ &  c, b\\
$R_{\rm rg}$ & RG radius    & $64\,R_{\odot}$   &  b   \\
$a$          & orbital separation  & 0.9\,au           &  c   \\
$i$          & orbital inclination & $55^{\circ}$      &  b   \\
\hline
\end{tabular}
\end{center}
(a) \cite{2004A&A...415..609S}; (b) \cite{2025ApJ...983...76H}; (c) \cite{1998MNRAS.296...77B}.
\end{table}

In our simulations, the computational domain is defined on a Cartesian grid, with the WD located at the origin of the coordinate system. The companion RG is modeled as an impenetrable body (through the definition of appropriate internal boundary conditions) positioned along the positive $x$-axis at a distance corresponding to the orbital separation. The explosion is initialized at the location of the WD as a spherical Sedov-type blast wave, characterized by an energy $E_{\rm bw}$ and an ejecta mass $M_{\rm ej}$. To explore the sensitivity of the shock evolution to the explosion parameters, we considered a factor-of-ten range for each quantity, spanning from $10^{43}$ to $10^{44}$~erg for $E_{\rm bw}$, and from $10^{-7}$ to $10^{-6}$~\msun\ for $M_{\rm ej}$, consistent with published estimates for the ejected mass and explosion energy in recurrent novae (e.g., \citealt{2005ApJ...623..398Y}). The baseline values, $E_{\rm bw} = 3\times 10^{43}$~erg and $M_{\rm ej} = 3\times 10^{-7}$~\msun, were selected as the logarithmic midpoints of these ranges. This ensures that the baseline model is representative of the parameter space explored and allows us to quantify the dynamical impact of moderate variations in explosion properties around a physically motivated central case. 

The maximum velocity of expanding ejecta ranges between $3000$ and $7500$~km~s$^{-1}$, consistent with the fastest velocities measured in 1946 ($\approx 5000$~km~s$^{-1}$; \citealt{Selvelli+92}). The initial blast radius is set to $r_{0} = 10^{12}$~cm (corresponding to $\approx 30$~minutes from the explosion), which is significantly smaller than the binary separation (0.9\,au $= 1.35\times 10^{13}$~cm). This choice ensures numerical stability and takes advantage of the kinetic energy of the ejecta overwhelmingly dominating over any gravitational binding: until this time the blast is spherically-symmetric and as yet unaffected dynamically by gravitational effects or the anisotropic distribution of circumbinary material. 

Following the approach of \cite{2017MNRAS.464.5003O} and as described in Sect.~\ref{sec:cbm}, the CBM surrounding the binary system includes a spherical wind component and an EDE, introduced to account for the gravitational focusing and likely accumulation of circumstellar material in the orbital plane. Systems such as RS\,Oph and V745\,Sco have shown clear evidence of bipolar shock morphologies and asymmetric X-ray emission, which have been effectively modeled by including a dense, flattened circumstellar structure in the orbital plane (e.g., \citealt{2009A&A...493.1049O, 2017MNRAS.464.5003O}).

Observational constraints on the wind from the RG companion and the density of the EDE are discussed in Sect.~\ref{sec:cbm}. For our HD simulations, we considered a range of EDE parameters in analogy with other symbiotic recurrent novae for which detailed 3D HD simulations have been successful in reproducing observed features. 
Motivated by these previous modeling efforts, we investigated a range of EDE densities ($10^6$ to $10^8$~cm$^{-3}$), thicknesses (1.5 to $6 \times 10^{13}$~cm), and scale lengths (4 to $8\times10^{14}$~cm), consistent with values adopted in earlier 3D simulations of RS\,Oph and V745\,Sco aimed at reproducing observed asymmetries and early-time light curves. These parameter ranges were chosen to bracket plausible configurations in \tcrb\ (consistent with radio constraints on the CBM at the low density end; Sect.~\ref{sec:cbm}),
enabling us to test how variations in EDE structure affect the degree of collimation, shock dynamics, and resulting X-ray observables. This comparative modeling approach allows us to build a predictive framework, and to identify potential diagnostics that future high-resolution X-ray observations of \tcrb\ can use to constrain the CBM properties post-eruption.

   \begin{figure}
   \centering
   \includegraphics[width=0.47\textwidth]{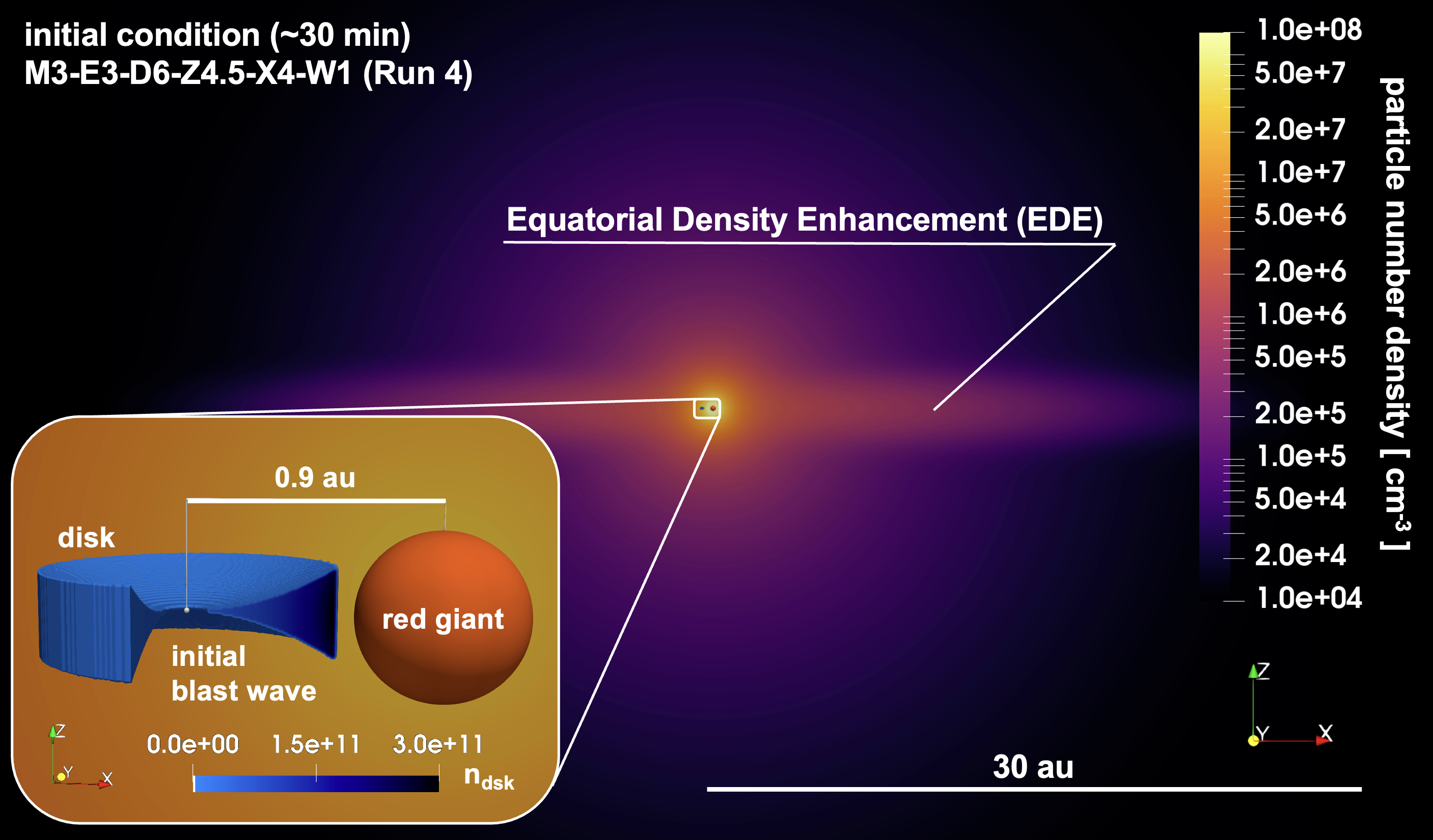}
   \caption{Colour-coded cross-section of the gas density distribution (in cm$^{-3}$) illustrating the initial conditions of model M3-E3-D6-Z4.5-X4-W1 (Run 4 in Table~\ref{tab2}). The cross-section emphasizes the EDE structure, visible as an enhanced density in the equatorial plane. The inset provides a zoomed view of the initial geometry of the \tcrb\ system: the WD and the spherical blast wave are located at the origin (white sphere, left), while the RG companion is positioned along the $x$-axis at $x = 0.9$ au (orange sphere, right). The accretion disk (blue) is centered on the WD.}
   \label{ini_cond}%
   \end{figure}

In addition, we included a flared, dense accretion disk around the WD, adopting the shape described by \citet{2000ApJ...534L.189H}. A similar disk structure was previously included in 3D HD models of U\,Sco \citep{2010ApJ...720L.195D}, but not in other 3D nova simulations such as those of RS\,Oph \citep{2009A&A...493.1049O}, V745\,Sco \citep{2017MNRAS.464.5003O}, or V407\,Cyg \citep{2012MNRAS.419.2329O}. According to \citet{2025A&A...694A..85P}, the tidal-truncation radius (which sets the upper limit for the disk’s outer radius) is $0.33 \pm 0.02$\,au (a similar value has been suggested by \citealt{2025A&A...701A.176M}). In our simulations, we adopted an extreme case with the disk extending out to $R_{\rm dsk} = 0.5$\,au to evaluate the maximal impact such a structure could have on the evolution of the blast wave\footnote{Even larger values for the disk radius have been proposed in the literature, see for example \citet{2024MNRAS.532.1421T}, who proposed a disk radius of $\approx 1$\,au.}. The disk's maximum vertical extent was set to $d_{\rm dsk} = 0.3$\,au at the outer edge, consistent with a flared geometry.

   \begin{figure}
   \centering
   \includegraphics[width=0.45\textwidth]{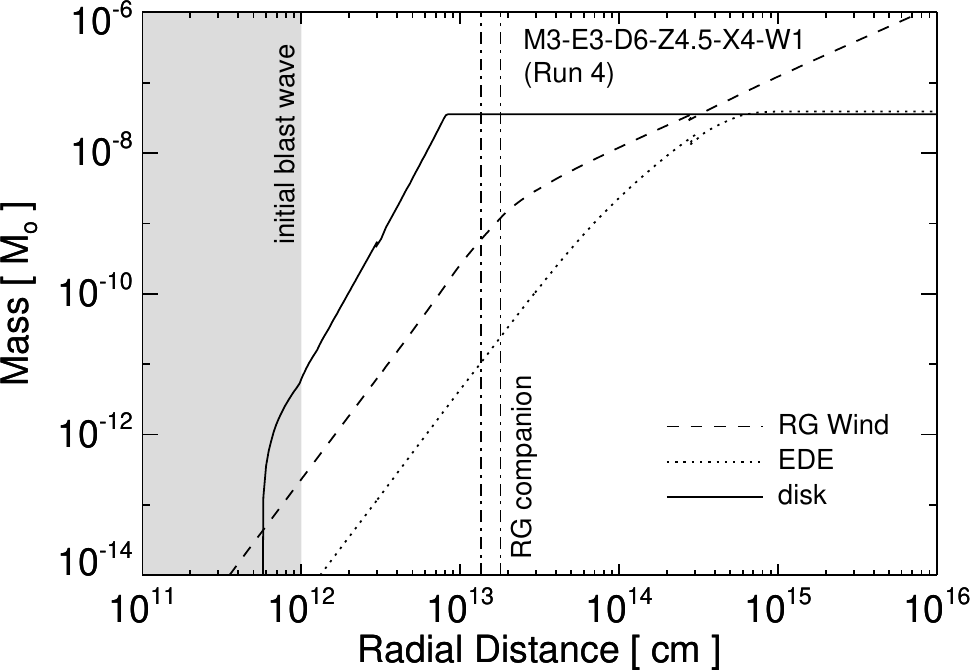}
   \caption{Cumulative mass of the CBM enclosed within a given radius as a function of radial distance from the WD, for model M3-E3-D6-Z4.5-X4-W1 (Run 4 in Table~\ref{tab2}). The plot shows contributions from the spherical RG wind (dashed line), the EDE (dotted line), and the accretion disk (solid line). The shaded region on the left marks the extent of the initial blast wave; the CBM is not described within this zone in the simulation. Vertical dot-dashed lines indicate the position of the RG companion along the $x$-axis.}
   \label{mass_profiles}%
   \end{figure}

To amplify the role of the disk in shaping the early interaction between the ejecta and the surrounding medium, we also assumed a high density contrast: the disk density was set to be three orders of magnitude higher than that of the ambient CBM at the same location\footnote{This effectively means that each portion of the disk is 1000 times denser than the surrounding medium would be if the disk were not present.}. This choice represents an extreme case that maximizes the disk’s contribution to the X-ray emission. The resulting total mass of the disk in this configuration is approximately $3 \times 10^{-8}\,M_\odot$. While this value likely exceeds the actual mass of the disk in \tcrb, it provides an upper bound for evaluating its potential observational signatures. Following the outburst, we plan to refine the disk parameters by comparing our models with observational data, aiming to constrain its geometry and density structure more precisely.

\begin{table*}
\caption{Explored parameters and initial conditions for the 3D HD models of the \tcrb\ outburst$^{(a)}$.}
\label{tab2}
\begin{center}
\begin{tabular}{llccccccc}
\hline
\hline
\# Run & Model$^{(b)}$  & \mej\  & \ebw\ & $n_{\rm ede}$ &  $h_z$ &  $h_x^{(c)}$  &  $n_{\rm w}$ & disk \\
    &   & [$10^{-7}$ \msun] & [$10^{43}$ erg] & [$10^{7}$ cm$^{-3}$] & [$10^{13}$ cm] & [$10^{14}$ cm]  &  [$10^{-3}$ cm$^{-3}$]  \\ \hline
1  & M3-E3-D6-Z4.5-X8-W5   & 3.0 & 3.0  &  0.1  &   4.5  &   8.0* &   5.0* & Y\\
2  & M3-E3-D6-Z4.5-X8-W1   & 3.0 & 3.0  &  0.1  &   4.5  &   8.0* &   1.0 & Y \\
3  & M3-E3-D8-Z4.5-X8-W1   & 3.0 & 3.0  &  10.*  &   4.5  &   8.0* &   1.0 & Y \\
\bf 4& \bf M3-E3-D6-Z4.5-X4-W1& \bf 3.0& \bf 3.0& \bf 0.1& \bf 4.5& \bf 4.0 & \bf 1.0 & \bf Y \\
5  & M3-E3-D7-Z3-X8-W1     & 3.0 & 3.0  &  1.0*  &   3.0*  &   8.0* &   1.0 & Y \\
6  & M3-E3-D7-Z1.5-X8-W1   & 3.0 & 3.0  &  1.0*  &   1.5*  &   8.0* &   1.0 & Y \\
7  & M3-E3-D7-Z6-X8-W1     & 3.0 & 3.0  &  1.0*  &   6.0*  &   8.0* &   1.0 & Y \\
8  & M3-E3-D6-Z3-X8-W1     & 3.0 & 3.0  &  0.1  &   3.0*  &   8.0* &   1.0 & Y \\
9  & M3-E3-D8-Z3-X8-W1     & 3.0 & 3.0  &  10.*  &   3.0*  &   8.0* &   1.0 & Y \\
10 & M3-E10-D7-Z3-X8-W1    & 3.0 & 10.*  &  1.0*  &   3.0*  &   8.0* &   1.0 & Y \\
11 & M3-E1-D7-Z3-X8-W1     & 3.0 & 1.0*  &  1.0*  &   3.0*  &   8.0* &   1.0 & Y \\
12 & M10-E3-D7-Z3-X8-W1    & 10.* & 3.0  &  1.0*  &   3.0*  &   8.0* &   1.0 & Y \\
13 & M1-E3-D7-Z3-X8-W1    & 1.* & 3.0  &  1.0*  &   3.0*  &   8.0* &   1.0 & Y \\
14 & M3-E3-D6-Z4.5-X4-W1-NOD&3.0 & 3.0  &  0.1  &   4.5  &   4.0 &   1.0 & N*\\
\hline
\end{tabular}
\end{center}
(a) An asterisk next to a value indicates that the corresponding parameter has been modified relative to the reference model, Run 4 (M3-E3-D6-Z4.5-X4-W1), which is highlighted in bold. All models assume an initial blast wave radius of $r_{0} = 10^{12}$~cm ($\approx 14\,R_{\odot}$). 
(b) Named according to: {\em M}, ejecta mass; {\rm E}, explosion energy; {\em D}, peak hydrogen logarithmic number density of the EDE; {\em Z}: EDE $z$ scale height; {\em X}, EDE $x$ and $y$ scale length; {\em W}, stellar wind hydrogen number density at 1 pc; {\em NOD}, absence of accretion disk; see Sect.~\ref{sec:model} for details.
(c) We assumed $h_y = h_x$, consistent with an axisymmetric EDE.
\end{table*}

Figure~\ref{mass_profiles} presents the total mass of the CBM enclosed within a given radius as a function of distance from the WD, for one of the models analyzed (M3-E3-D6-Z4.5-X4-W1, Run 4 in Table~\ref{tab2}). The plot shows that the majority of the shocked CBM within a radius of approximately $2 \times 10^{14}$~cm ($\sim 13$~au) originates from the disk. At larger distances, the RG spherical wind becomes the dominant component. Interestingly, the EDE contributes only marginally across all spatial scales. However, as discussed in Sect.~\ref{sec:lightcurves}, the EDE is the primary driver shaping both the bipolar morphology of the nova remnant and its X-ray emission characteristics after the first few days of eruption (the accretion disk dominates at the earliest times).

We explore a broad parameter space by varying $E_{\rm bw}$ and $M_{\rm ej}$, and the properties of the surrounding CBM, including the density and geometry of both the EDE and the accretion disk. Table~\ref{tab2} summarizes the models analyzed here. The naming convention for the models is as follows: {\em M} denotes the ejecta mass in units of $10^{-7}$ \msun; {\em E} the explosion energy in units of $10^{43}$ erg; {\em D} the peak hydrogen number density of the EDE (logarithmic scale); {\em Z} the scale height of the EDE in the $z$ direction in units of $10^{13}$ cm; {\em X} the scale length of the EDE in the $x$ and $y$ directions in units of $10^{14}$ cm; {\em W} the hydrogen number density of the stellar wind at 1 pc; {\em NOD} denotes the absence of an accretion disk. 

   \begin{figure*}
   \centering
   \includegraphics[width=\textwidth]{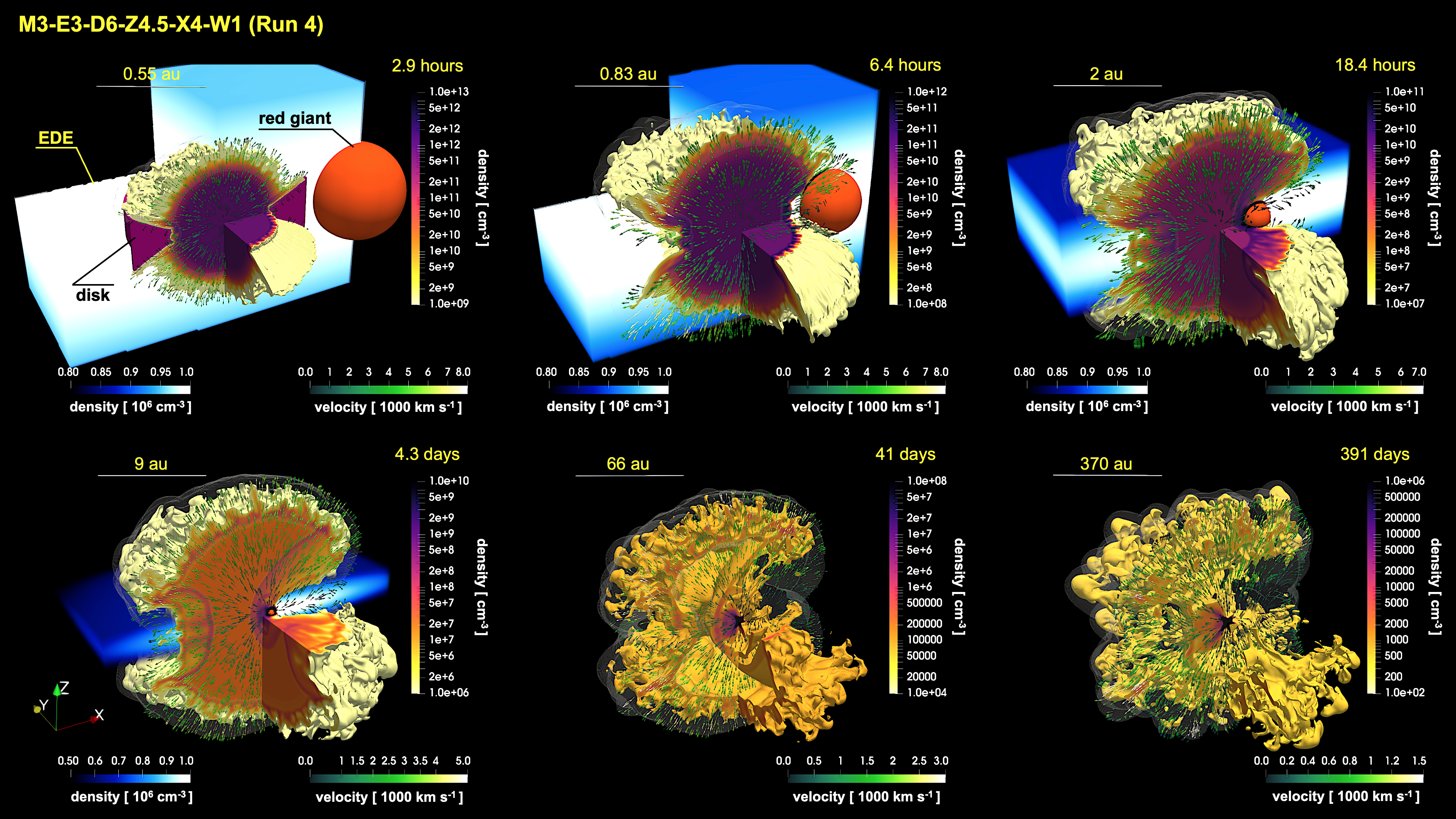}
   \caption{Evolution of the blast wave in reference model M3-E3-D6-Z4.5-X4-W1 (see Run 4 in Table~\ref{tab2} for model parameters). Each panel displays the 3D density structure of the nova remnant at the indicated time (upper right corner of each panel), with a yardstick in the upper left corner denoting the spatial scale. The colored isosurface traces the distribution of the ejecta, with color representing local gas density (see color bar at right of each panel). The isosurface is partially clipped to expose the internal structure of the remnant. A semi-transparent gray isosurface outlines the forward shock. The RG companion is shown as an orange sphere, most visible at early times (top panels). The accretion disk, rendered in violet around the WD, appears only in the earliest frame (upper left panel). Green arrows indicate the velocity field of the outflowing plasma, with arrow color encoding the flow speed (color bar at lower right). The EDE is visible at times earlier than 40 days (upper panels and lower left panel), rendered in blue (color bar at lower left), and is partially clipped to reveal the remnant structure. The blast wave is visibly collimated along the polar directions, shaped by the combined influence of the EDE and the accretion disk.}
   \label{evol}%
   \end{figure*}

\section{Results}
\label{sec:results}

\subsection{Hydrodynamic evolution of the outburst}
\label{sec:hydro}

The HD evolution of a nova outburst in a symbiotic-like system such as \tcrb\ is shaped by the complex interaction between the blast wave and the dense, anisotropic CBM. This environment strongly influences the remnant’s morphology and thermal structure, leading to pronounced emission asymmetries (e.g., see \citealt{2009A&A...493.1049O, 2017MNRAS.464.5003O}). Our 3D simulations show that the thermonuclear runaway on the WD surface triggers a rapidly expanding blast wave and ejecta, which evolve through distinct phases shaped by system geometry and local density gradients. Figure~\ref{evol} illustrates the evolution for model M3-E3-D6-Z4.5-X4-W1 (Run 4 in Table~\ref{tab2}), which we adopt as our reference case. The other models exhibit qualitatively similar behavior.

   \begin{figure*}
   \centering
   \includegraphics[width=0.93\textwidth]{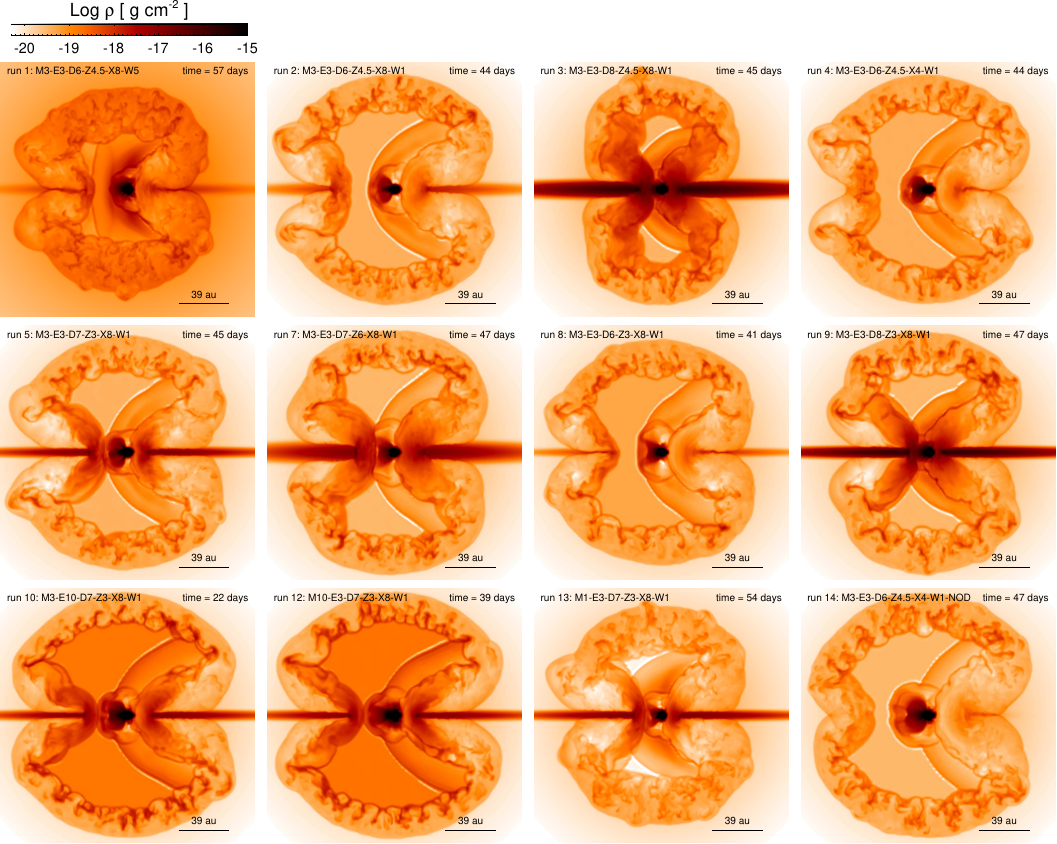}
   \caption{The 2D cross-sections in the $(x, z)$ plane display the logarithmic gas density distribution of the nova remnant at the indicated times for most models listed in Table~\ref{tab2} (the model name is shown in the upper left corner of each panel). Runs 6 and 11 are omitted since their morphology is similar to Run 5. The bipolar structure of the blast wave is approximately aligned with the $z$-axis. A scale bar in the lower right corner of each panel indicates the physical length scale. The EDE itself is not visible in models M3-E3-D6-Z4.5-X4-W1 (Run 4) and M3-E3-D6-Z4.5-X4-W1-NOD (Run 14), as they adopt the most compact EDE configuration among all models considered (see Table~\ref{tab2}), and the entire EDE has already been shocked by the time shown.}
   \label{models}%
   \end{figure*}

At the moment of outburst, a spherical blast wave is launched from the surface of the WD. This initial shock front propagates supersonically into the surrounding medium with typical velocities exceeding several thousand km~s$^{-1}$. The immediate environment is shaped by the presence of the accretion disk around the WD and the extended wind of the RG companion. Depending on the disk's size and density profile, it can be rapidly ablated, shocked, and partially entrained into the ejecta flow within just a few hours of the outburst (see upper left panel in Fig.~\ref{evol}). However, in some symbiotic novae (as, for instance, RS\,Oph and V407\,Cyg) the accretion disk is often flattened or disrupted prior to the outburst by interactions with the dense RG wind (e.g. \citealt{2008A&A...484L...9W, 2015ApJ...806...27P}), and thus its role in the shock dynamics can be secondary compared to the effect of the EDE. To account for this scenario, we expanded our suite of simulations by introducing model M3‑E3‑D6‑Z4.5‑X4‑W1‑NOD (Run 14; see Table~\ref{tab2}). This model shares the same setup as our reference model (Run 4, M3‑E3‑D6‑Z4.5‑X4‑W1) but excludes the accretion disk. By directly comparing the two models (e.g., see Sect.~\ref{sec:lightcurves} and Fig.~\ref{lightcurve}), we can quantify the disk’s impact on the outflow dynamics and observable properties, thereby better constraining its role in shaping the morphology and energetics of the nova remnant.

A few hours after the outburst, the blast wave encounters the RG companion, which has a radius of approximately 64~$R_{\odot}$ and lies at a distance of 0.9~au from the WD (see Table~\ref{tab1}). The companion star acts as a physical obstacle to the expanding shock, partially shielding the blast and causing the shock front to refract around its surface. This interaction leads to the formation of a bow shock on the leading side of the star, which heats the surrounding ejecta and CBM. On the trailing side, the convergence and reflection of the shock produce a hot, dense wake (see upper panels in Fig.~\ref{evol} and Fig.~\ref{models} for most of the models listed in Table~\ref{tab2}; see also discussion in \citealt{2017MNRAS.464.5003O}). Moreover, the geometry of the bow shock induces a secondary collimation of the ejecta along the polar axis due to HD focusing effects (see also the lower panels in Fig.~\ref{evol}). However, this collimation remains modest compared to the stronger, primary collimation produced by the accretion disk and the EDE. Together, these processes significantly shape the morphology of the remnant and may leave observable imprints in the emission line profiles.

A critical feature that dominates the evolution of the blast is the presence of the EDE. As mentioned in Sect.~\ref{sec:model}, this disk-like structure forms via the gravitational focusing and accumulation of the RG wind material in the orbital plane. The EDE introduces a substantial anisotropy in the ambient medium, with densities several orders of magnitude higher in the equatorial regions than in the polar directions. As the blast wave expands, it encounters this dense equatorial torus, leading to (see Fig.~\ref{evol}): i) strong shock deceleration in the equatorial plane due to the high density; ii) shock reflection and rarefaction phenomena at the interface between the expanding ejecta and the EDE; iii) collimation of both blast and ejecta into the polar directions, resulting in a distinctly bipolar morphology (e.g., \citealt{2017MNRAS.464.5003O}). This collimation effect is present in all the models analyzed (see Fig.~\ref{models}).

The degree of collimation of the ejecta is primarily governed by the explosion energy and by both the density and vertical extent of the EDE (see Fig.~\ref{models}). As shown in the lower panels of Fig.~\ref{evol}, for the representative parameters of the nova explosion explored in our study (namely explosion energies of a few $\times 10^{43}$~erg and ejecta masses of a few $\times 10^{-7}$~M$_{\odot}$) the blast wave expands significantly faster along the polar directions than in the equatorial plane (see also Fig.~\ref{models} for most of the models analyzed in this study). This anisotropic expansion is driven by the lower density environment above and below the orbital plane, in contrast to the denser EDE material that hinders equatorial propagation. The resulting shock morphology resembles a prolate ellipsoid or a twin-lobed structure, consistent with findings in previous numerical studies (e.g., \citealt{2009A&A...493.1049O, 2010ApJ...720L.195D, 2017MNRAS.464.5003O}). This collimation effect is evident across all our models within the parameter space probed for the EDE (see Fig.~\ref{models}). 

Notably, even modest increases in the EDE density or a reduced vertical extent can produce pronounced asymmetries in the remnant structure (even without the presence of an accretion disk), potentially giving rise to the bipolar or axisymmetric morphologies observed in some nova remnants. By the end of the simulation (at an age of approximately one year), after the EDE is fully shocked, the blast wave propagates through the spherically symmetric wind of the RG companion. At this stage, the ejecta continue to expand in all directions, gradually approaching spherical symmetry over time. This evolution mirrors that observed in SN blast waves, where the ejecta initially encounter a highly asymmetric environment but tend to become more spherical as they expand into a more isotropic, spherically symmetric medium (e.g., \citealt{2021A&A...654A.167U, 2025A&A...699A.305O}).

As the blast interacts with the disk and the EDE, a reverse shock forms, propagating inward through the ejecta (see all panels except the last in Fig.~\ref{evol}; see also Fig.~\ref{models}). This reverse shock compresses and heats the inner ejecta to temperatures of several million Kelvin. The resulting EM distributions as a function of electron temperature ($kT_{\rm e}$) and ionization parameter ($\tau$) display distinctive features, including multiple peaks associated with plasma components at temperatures well above 0.1~keV, and significant departures from collisional ionization equilibrium (CIE), particularly at late times (see Fig.~\ref{em_dist} and discussion in Appendix~\ref{sec:em_dist}). While the forward shock predominantly heats CBM, the reverse shock contributes significantly to the high-energy emission from the shocked ejecta. According to the EM distributions (see lower panels in Fig.~\ref{em_dist}), the shocked ejecta can become a major source of X-ray line emission at lower energies (e.g., O~VIII, Fe~XVII), while the hotter shocked CBM is expected to dominate higher energy lines (e.g., Si~XV, Fe~XXV; see also discussion in Sect.~\ref{sec:lightcurves} and \ref{sec:spectra}). The evolution of the EM distribution throughout the outburst is discussed in detail in Appendix~\ref{sec:em_dist}.

At later stages (more than two weeks post-outburst), the remnant develops a morphology featuring two opposing polar cavities filled with hot, shocked plasma, both enclosed by a common central denser equatorial shell (see lower panels of Fig.~\ref{evol} and Fig.~\ref{models}). Shocks propagating both polewards and along the equatorial plane are expected to contribute to the X-ray emission, though with different spectral characteristics due to the anisotropic density distribution of the CBM. In the polar regions, the ejecta encounter lower densities, experience less deceleration, and maintain higher velocities. Since for a strong shock the temperature of the shocked gas scales with the square of the shock speed, this most likely results in harder X-ray emission. Conversely, the interaction with the denser equatorial material is expected to generate softer X-ray emission and can lead to asymmetric absorption of redshifted line components along the line-of-sight (LoS). This results in blueshifted, asymmetric line profiles, often marked by enhanced blue wings and suppressed red emission, as was first observed in the 2006 explosion of RS~Oph (\citealt{2009A&A...493.1049O, 2009ApJ...691..418D}; see also \citealt{2010ApJ...720L.195D, 2017MNRAS.464.5003O}).

\subsection{X-ray morphology and temporal evolution}
\label{sec:lightcurves}

   \begin{figure}[!t]
   \centering
      \includegraphics[width=0.43\textwidth]{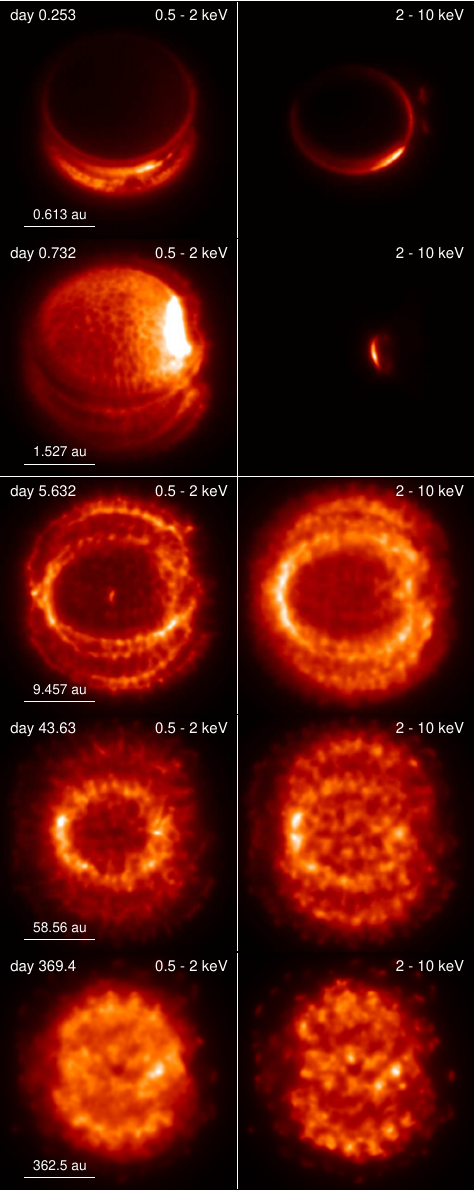}
   \caption{Synthetic X-ray emission maps from model M3-E3-D6-Z4.5-X4-W1 (Run 4) in the $[0.5, 2]$~keV (left panels) and $[2, 10]$~keV (right panels) energy bands at the indicated times (top to bottom). Each image is normalized to its respective maximum intensity for visualization purposes. The scale bar in the lower-left corner of each panel denotes the physical length scale; at the distance of \tcrb\ (890~pc), a length of 400~au corresponds to $0.45^{\prime\prime}$.}
   \label{emission_maps}%
   \end{figure}
   
As described in Appendix~\ref{app:synthesis}, we synthesized the X-ray emission predicted by our HD models, accounting for contributions from the shocked ejecta and the CBM, including the accretion disk and the EDE. The synthesis accounts for Doppler shifts and line broadening due to the LoS component of the plasma velocity, photoelectric absorption by the interstellar medium, CBM, and ejecta, as well as deviations from both electron-proton temperature equilibration and ionization equilibrium. To reproduce the observed inclination of the \tcrb\ system \citep{2025ApJ...983...76H}, we rotated the system by $55^\circ$ about the $x$-axis (see Table~\ref{tab1}), following a prior $20^\circ$ rotation about the $z$-axis to represent a generic orbital phase at outburst (this value may vary with the actual eruption time). The observer’s LoS is assumed to be along the negative $y$-axis, enabling direct comparison with observations. Figure~\ref{emission_maps} presents a sequence of synthetic X-ray emission maps in the $[0.5, 2]$~keV (left panels) and $[2, 10]$~keV (right panels) energy bands for our reference model, shown at the labeled times (from top to bottom), illustrating the remnant’s morphological evolution.

In the earliest stages (within the first few hours after the outburst), most of the observed emission originates from the blast wave propagating through the accretion disk on the front side of the remnant. In contrast, X-ray emission from the rear side is heavily absorbed by the optically thick ejecta along the LoS (see first row in the figure). Around day one, the dominant X-ray–emitting feature becomes the bright reflected shock generated as the blast wave collides with the RG companion (second row). At later times, the remnant’s morphology evolves into a more complex structure. Initially, this is characterized by a bright, ring-like feature resulting from the forward shock expanding into the dense EDE and the reverse shock propagating back into the ejecta (third row). As the remnant continues to evolve, its morphology changes further. By about one month after the outburst, a distinctly bipolar structure emerges, which is especially pronounced in the hard X-ray band (see last two rows in the figure). Finally the remnant becomes increasingly isotropic (see discussion in Sect.~\ref{sec:hydro}, and \citealt{2021A&A...654A.167U}). 

It is worth noting that after one year of evolution (lower panels in Fig.~\ref{emission_maps}), the nova remnant is expected to reach a size of roughly 800~au, corresponding to an angular diameter of approximately $\approx 0.9^{\prime\prime}$ at the distance of \tcrb. This angular size lies below the spatial resolution limits of current X-ray observatories such as XMM-Newton and XRISM. The Chandra observatory, with its superior angular resolution of $\sim 0.5^{\prime\prime}$, could in principle resolve the remnant in a few pixels at later stages (after day 40) and detect asymmetries in the shock structure, particularly in deep exposures. However, given the small angular extent and potentially limited photon statistics at late times, the remnant may remain marginally resolved in practice.

Nonetheless, the synthetic X-ray maps remain valuable for identifying which remnant structures dominate the emission. They clearly demonstrate how the CBM (including the EDE, the accretion disk, and the RG companion) critically influences both the thermal properties and the observable morphology of the X-ray–emitting plasma over time. These results underscore the importance of multi-band X-ray observations for probing the geometry and physical conditions of nova remnants like \tcrb, and highlight the essential role of 3D HD simulations in interpreting the data.

   \begin{figure*}
   \centering
   \includegraphics[width=0.9\textwidth]{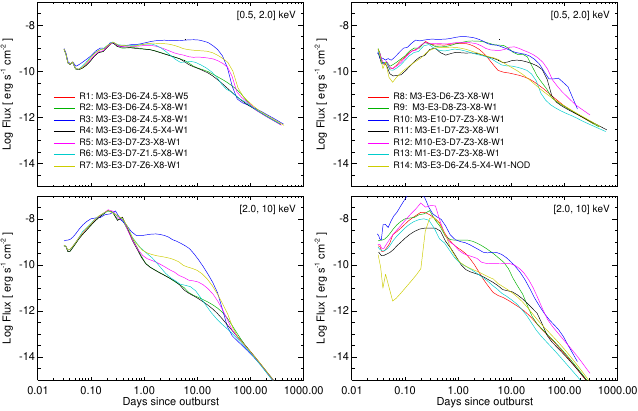}
   \caption{X-ray lightcurves synthesized from all the HD models listed in Table~\ref{tab2}, in the $[0.5,2]$~keV (upper panels) and $[2,10]$~keV (lower panels) bands. 
   }
   \label{lightcurve_all}%
   \end{figure*}

   \begin{figure*}
   \centering
   \includegraphics[width=0.9\textwidth]{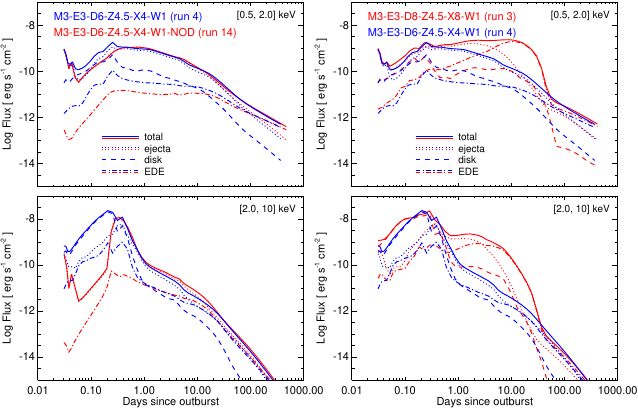}
   \caption{Same as in Fig.~\ref{lightcurve_all} for some selected models in Table~\ref{tab2}. Left panels: comparison between two models that differ only by the presence (Run 4, blue solid line) or absence (Run 14, red solid line) of the accretion disk. The dotted, dashed, and dot-dashed lines indicate the contributions to the total emission from the shocked ejecta, the shocked disk material, and the shocked plasma from the CBM (EDE and RG wind), respectively. Right panels: comparison between two models differing in the geometry and density of the EDE (Runs~3 and 4).}
   \label{lightcurve}%
   \end{figure*}

Figure~\ref{lightcurve_all} shows the resulting X-ray lightcurves for all models listed in Table~\ref{tab2}, in the $[0.5, 2]$~keV (upper panels) and $[2, 10]$~keV (lower panels) energy bands. Overall, the models exhibit a qualitatively similar evolution: an initial rapid rise in X-ray flux during the first few hours after the outburst (particularly pronounced in the hard band) followed by a more gradual decline over subsequent months as the remnant expands and cools. The left panels of Fig.~\ref{lightcurve} explore this behavior in greater detail for the reference model M3-E3-D6-Z4.5-X4-W1 (Run 4; blue solid line). Here, the X-ray lightcurves are decomposed into the contributions from the distinct shocked plasma components: the ejecta, the disk material, and the CBM (including EDE and RG wind). These panels also include a direct comparison with an identical model lacking an accretion disk (Run 14, M3-E3-D6-Z4.5-X4-W1-NOD; red lines) to highlight the impact of the disk on the observable X-ray emission.

Figure~\ref{lightcurve} demonstrates that the most significant differences between the two models (Runs 4 vs. 14) arise within the first $\sim 8$ hours of evolution, particularly in the hard X-ray band (left panels in the figure). During this early phase, the soft X-ray emission is produced in roughly equal measure by the shocked ejecta and the shocked disk material (see dotted and dashed blue lines in the upper left panel of Fig.~\ref{lightcurve}). By contrast, the hard X-ray emission is almost entirely dominated by the shocked disk material in our reference model (dashed blue line in the lower left panel of Fig.~\ref{lightcurve}), which is responsible for the pronounced emission peak around 5 hours after the outburst. In both energy bands, the contribution from the shocked CBM remains negligible at these early times, as the blast wave has not yet swept up enough material in the dense equatorial regions to rival the other components. This explains the stark difference between the two models: in the absence of the disk, the hard-band flux is suppressed by several orders of magnitude, since it lacks the additional high-temperature shocked plasma produced by the disk’s rapid ablation.

At later times (beyond $\sim 8$ hours), as the shock continues to expand into the CBM and lower-density regions, the overall X-ray flux gradually declines in both bands, and the relative differences between the models become less pronounced. The emission increasingly reflects the contribution of the shocked ejecta and CBM, while the role of the shocked disk material rapidly diminishes. These results highlight the critical influence of the accretion disk on the early X-ray observables of the outburst, particularly in the hard band, and demonstrate how X-ray observations within the first hours after eruption can provide powerful diagnostics of the disk’s presence and, possibly, of its density and structure. Exposure times on the order of $10-50$\,ks (corresponding to the first $\approx 3-14$ hours of evolution) would be sufficient to detect the enhanced hard X-ray emission produced by the shocked disk and to distinguish models with and without a disk (see Sect.~\ref{sec:spectra}). Given that the disk’s effect is more pronounced in the hard band (above 2 keV), XRISM/Resolve is better suited than XMM-Newton/RGS for this task, offering greater sensitivity in the relevant energy range.

After the peak of X-ray emission at $\approx 5$ hours after the explosion, all X-ray–emitting components show a gradual decline, though with different rates: the shocked disk material exhibits the steepest decline (as mentioned above), while the emission from the shocked CBM decreases more slowly. As a result, at later times (around 3 months after the outburst in models M3-E3-D6-Z4.5-X4-W1 and M3-E3-D6-Z4.5-X4-W1-NOD), the X-ray emission becomes dominated by the shocked material from the CBM, particularly in the soft X-ray band. This effect is less pronounced in the reference model, which assumes a relatively compact EDE confined to the equatorial plane (see Fig.~\ref{lightcurve_all}). However, in other models, the impact of the EDE can be much more significant, especially when the EDE extends further in the radial direction within the equatorial plane (as in models with $h_{\rm x} = h_{\rm y} = 8\times 10^{14}$\,cm; see Table~\ref{tab2}), has a larger vertical thickness (e.g., Run 7, M3-E3-D7-Z6-X8-W1), or possesses a higher density (as in Run 3, M3-E3-D8-Z4.5-X8-W1, and Run 9, M3-E3-D8-Z3-X8-W1). The ejecta mass and explosion energy also play an important role in shaping both the timing and relative contribution of different plasma components to the X-ray emission (see Fig.~\ref{lightcurve_2} and discussion in Sect.~\ref{sec:constr_mass_en}).

The effect of the EDE is clearly evident between day 1 and day 100 in both energy bands, as shown in Fig.~\ref{lightcurve_all}. Among the models considered, M3-E3-D8-Z4.5-X8-W1 (Run 3) displays the most pronounced influence of the EDE on the X-ray flux, in both the soft and hard bands (see right panels in Fig.~\ref{lightcurve}). Notably, in this model, the X-ray flux remains almost constant between day 1 and day 10, and the peak of the soft X-ray emission is delayed until around day 20. By comparing Runs 3 and 4, we note that, increasing the EDE density by two orders of magnitude results in a larger amount of shocked CBM, which contributes more significantly to the X-ray emission. As a consequence, the CBM-dominated phase appears earlier (within the first day after the eruption) compared to our reference model (Run 4). The higher density of the EDE also leads to lower post-shock temperatures, enhancing the emission in the soft X-ray band. Meanwhile, the broader equatorial extent of the EDE causes a pronounced peak in the X-ray emission around 20–30 days after the eruption. After day~50, the differences between the two models (Runs~3 and 4) become negligible, and their lightcurves converge.

From the fitting of synthetic spectra (see Sect.~\ref{sec:spectra}), we estimated that the statistical uncertainty in the determination of fluxes is below ~1\% during the first 20 days of evolution. This suggests that, in principle, it is possible to discriminate between different models during the early phases, when the effects of the EDE are expected to be most pronounced. However, we note that instrumental calibration uncertainties are typically at the level of $\sim 10$\%, and additional systematics (arising from atomic data, elemental abundances, and plasma modeling assumptions) can contribute further uncertainty. These factors will inevitably affect the inference drawn from model–data comparisons. 

   \begin{figure*}[!t]
   \centering
      \includegraphics[width=0.7\textwidth]{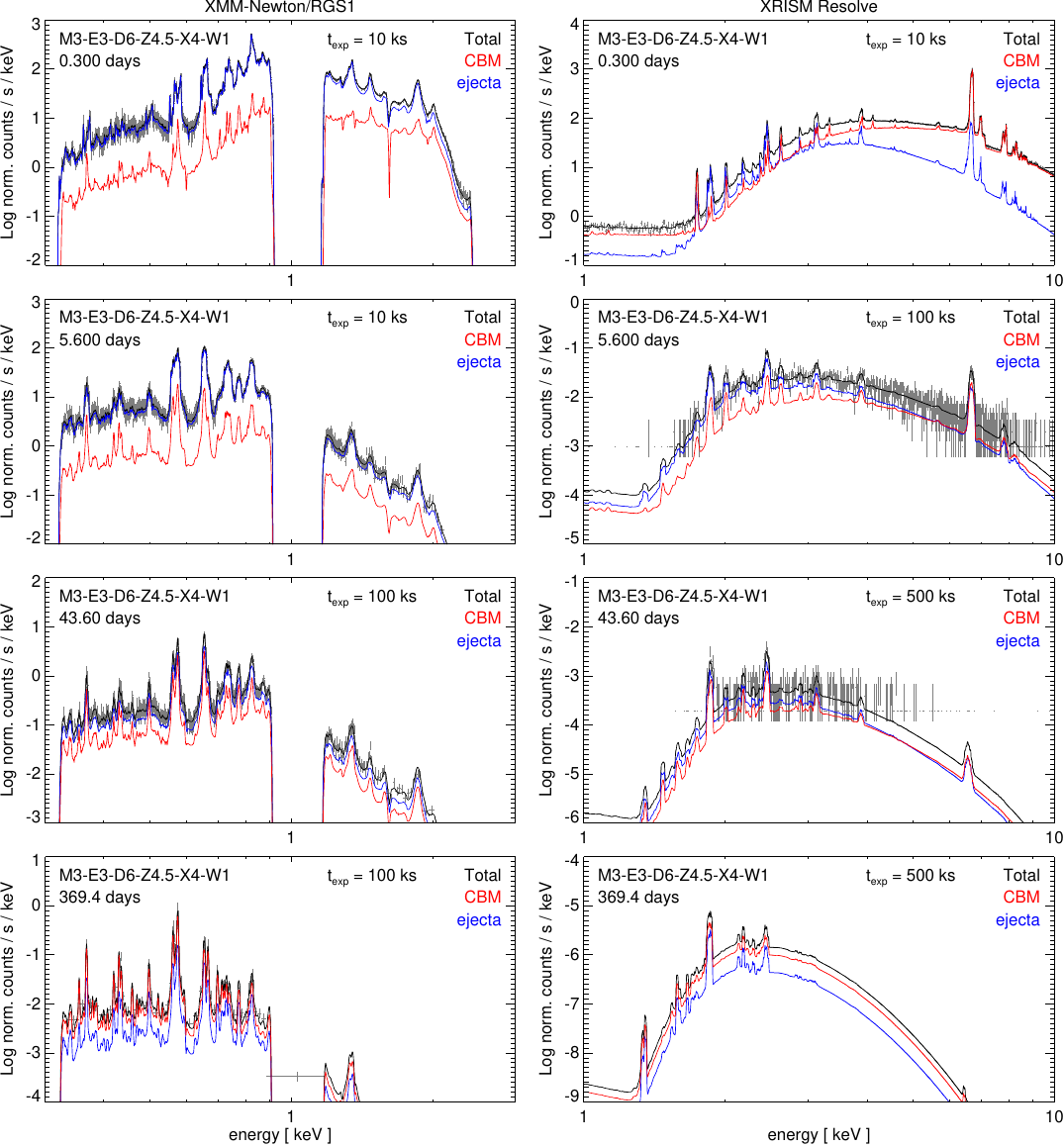}
   \caption{Full XMM-Newton/RGS1 (left panels) and XRISM/Resolve (right panels) synthetic spectra derived from model M3-E3-D6-Z4.5-X4-W1 (Run 4) at the indicated epochs. The simulated observed spectra are shown as gray crosses and were computed assuming an exposure time of 10~ks during the first day (top row) and between 10 and 500~ks at later epochs, depending on the signal-to-noise ratio. Note that the gap in the XMM-Newton/RGS1 data around 1~keV is due to the failure of one of the RGS1 CCDs, which occurred very early in the mission. In addition, the XRISM/Resolve spectrum at day~369 (bottom right panel) shows no detectable signal, even with an exposure time of 500~ks. For comparison, the ideal high-resolution synthetic spectra are overplotted as black lines. The figure also shows the separate contributions from shocked CBM (including disk, EDE, and RG wind; red) and shocked ejecta (blue), highlighting how each component shapes the overall X-ray emission.}
   \label{tot_spectra}%
   \end{figure*}

Despite these limitations, our results underscore the crucial role played by the density, geometry, and spatial extent of the EDE in shaping the late-time X-ray emission of the nova remnant. Our modelling of inter-eruption radio observations (Sect.~\ref{sec:cbm}) favors a lower density EDE in \tcrb\ (``D6" models: Runs 1, 2, 4, 8, and 14 in Fig.~\ref{lightcurve_all}), minimizing the EDE bump around day 20--30. X-ray observations of the imminent eruption of \tcrb\ will test this prediction and further constrain the properties of the CBM.

\subsection{X-ray spectra and line diagnostics}
\label{sec:spectra}

In this section, we explore the diagnostic potential of current X-ray missions, specifically assessing the capabilities of XMM-Newton/RGS and XRISM/Resolve to probe the physical properties of the X-ray emitting plasma. To this end, we synthesized spectra from these instruments based on our HD models of \tcrb’s outburst, sampling several representative epochs after the expected eruption: shortly after outburst (e.g., within a few hours), at a few days, around one month, and approximately one year later. For these synthetic observations, we adopted an exposure time of 10 ks during the first day of evolution, increasing to 100 ks for XMM-Newton/RGS and 500 ks for XRISM/Resolve at epochs beyond a few days, to capture both the rapid early evolution and the longer-term changes in the shocked plasma. We did not include synthetic Chandra/HETG spectra, primarily due to severe molecular contamination on the ACIS detector, which has significantly reduced its effective area in the soft X-ray band, precisely where \tcrb’s emission is expected to peak (see Appendix~\ref{app:synthesis}).

The resulting synthetic spectra cover the approximate energy ranges of $[0.3, 2.5]$~keV for XMM-Newton/RGS and $[1.5, 10]$~keV for XRISM/Resolve (the low energy cutoff being due to the gate valve anomaly; see Appendix \ref{app:synthesis}). These ranges roughly correspond to the two bands considered to derive the lightcurves presented in Sect.~\ref{sec:lightcurves}. Examples of spectra synthesized from our reference model (M3-E3-D6-Z4.5-X4-W1) are shown in Fig.~\ref{tot_spectra}.

The synthetic spectra show a gradual evolution of the X-ray emission over time. From $\approx 0.3$ to 369 days after the eruption the overall X-ray flux steadily decreases, with a more pronounced decline at higher energies (above 1~keV). Consequently, the source becomes progressively softer, consistent with the lightcurve behavior discussed in Sect.~\ref{sec:lightcurves}. For our reference model, the first month after the outburst is characterized by emission dominated by shocked ejecta (blue component in Fig.~\ref{tot_spectra}) at energies below 3~keV, while at higher energies (detected by XRISM/Resolve) the emission initially comes mainly from the shocked CBM (including disk, EDE, and RG wind; red component). This is followed by a phase where both components contribute comparably at energies above 3~keV (second and third rows in the figure). At later times, the shocked CBM becomes the dominant source of X-rays across the entire energy range (bottom row), reflecting the increasing amount of CBM material swept up by the forward shock, again consistent with the trends seen in the lightcurves.

We note that XMM-Newton will be able to detect the source with good signal-to-noise ratio throughout the entire simulated evolution ($\approx 1$\,year) using exposure times $\leq 100$~ks. In contrast, XRISM/Resolve will achieve good-quality spectra only during the first month after the eruption. By day~44, the spectra become very poor even with an exposure time of 500~ks, and at later times XRISM/Resolve is unlikely to detect any signal, even with similarly long exposures. However, if the gate valve, which has a beryllium window that attenuates the softest X-rays, is opened by the time of the \tcrb\ eruption, XRISM/Resolve could provide valuable observations below 2\,keV (down to 0.3~keV and potentially at higher energies as well) thanks to the significantly higher effective area at lower energies that would then be available.

   \begin{figure*}[!t]
   \centering
      \includegraphics[width=0.95\textwidth]{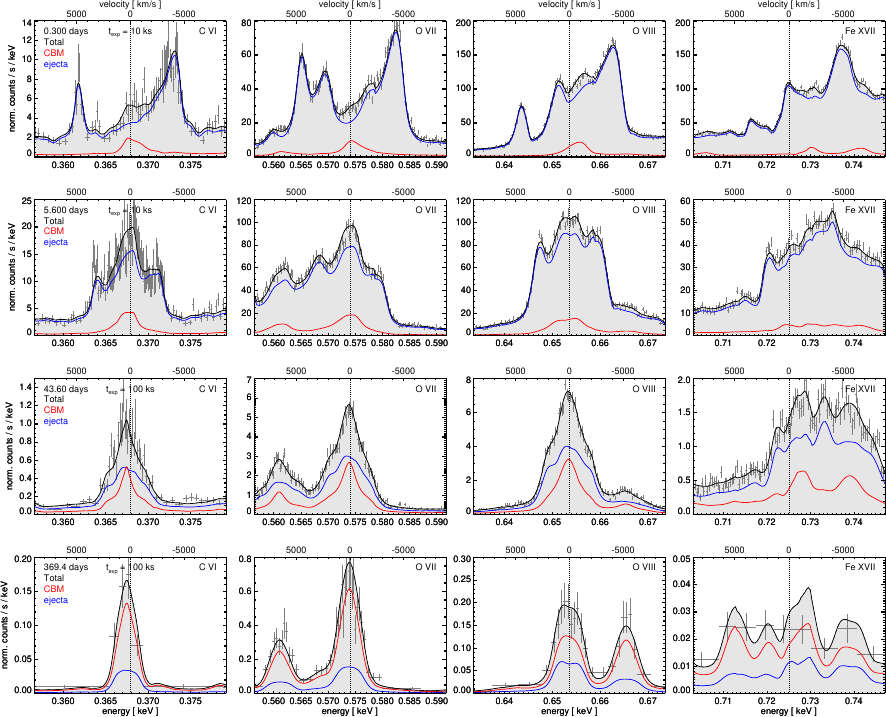}
   \caption{XMM-Newton/RGS1 synthetic spectra derived from model M3-E3-D6-Z4.5-X4-W1 (Run 4) at four representative epochs following the outburst (from top to bottom): approximately 7 hours, 6 days, 44 days, and 1 year. The figure shows close-up views of selected emission lines: C\,VI (0.37 keV), O\,VII (0.57 keV), O\,VIII (0.65 keV), and Fe\,XVII (0.72 keV). In each panel, the simulated observed spectra (gray crosses) were computed assuming exposure times of 10 ks during the early phase (top two rows) and 100 ks at later epochs (bottom two rows). Overlaid are the corresponding ideal synthetic spectra (black lines), together with the separate contributions from the shocked CBM (including disk, EDE, and RG wind; red) and the shocked ejecta (blue). The top axis in each panel indicates the Doppler velocity shifts relative to the rest-frame wavelength of the corresponding emission line.}
   \label{xmm_lines}%
   \end{figure*}

\begin{figure*}
   \centering
   \includegraphics[width=0.95\textwidth]{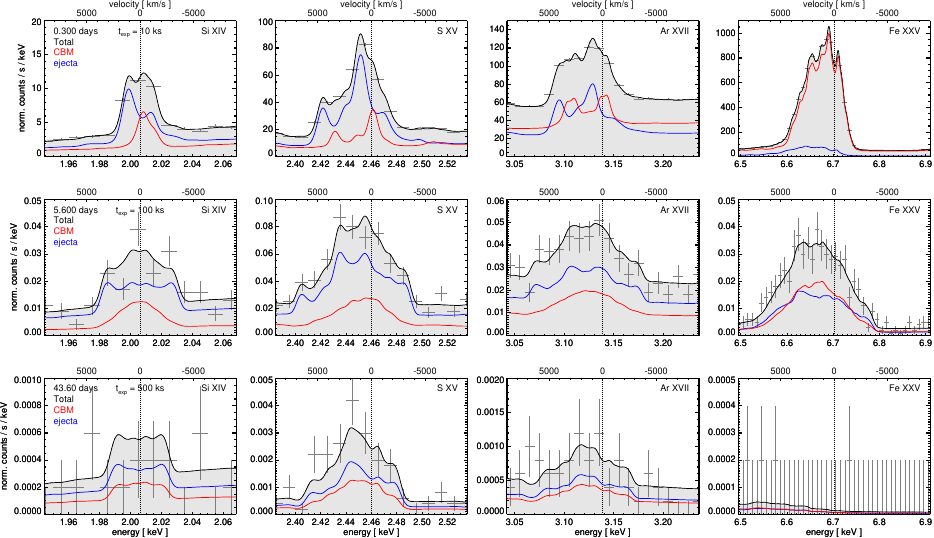}
   \caption{Same as Fig.~\ref{xmm_lines}, but showing selected emission lines as they would be observed with XRISM/Resolve. The lines include Si\,XIV (2.1 keV), S\,XV (2.46 keV), Ar\,XVII (3.14 keV), and Fe\,XXV (6.7 keV). Spectra are shown only for the first three epochs in Fig.~\ref{xmm_lines}, as the signal becomes negligible at later times.}
   \label{xrism_lines}%
   \end{figure*}

It is important to note that this evolution and the signal-to-noise of the spectra depend on the characteristics of the outburst (ejecta mass and explosion energy) and on the structure of the CBM. For instance, as shown in Appendix~\ref{app:add_spec}, the alternative model M3-E3-D8-Z4.5-X8-W1 (Run~3 in Table~\ref{tab2}) predicts an earlier dominance of shocked CBM emission, appearing just a few days after the eruption (see Fig.~\ref{lines_run3}), and good-quality XRISM/Resolve spectra even one month after the eruption (see Fig.~\ref{lines_xrism_run3}). This highlights how changes in the physical parameters (especially the geometry and density distribution of the CBM) can significantly affect the relative contributions of shocked ejecta and CBM, making the first month of X-ray emission a sensitive diagnostic of the circumbinary environment.

In our analysis, we focused on a set of emission lines that may serve as sensitive diagnostics of the shocked ejecta and CBM. In particular, we examined soft X-ray lines observable with XMM-Newton/RGS, such as C\,VI (0.37 keV), O\,VII (0.57 keV), O\,VIII (0.65 keV), and Fe\,XVII (0.72 keV), as well as higher-energy lines detectable by XRISM/Resolve, including Si\,XIV (2.1 keV), S\,XV (2.46 keV), Ar\,XVII (3.14 keV), and Fe\,XXV (6.7 keV). Figure~\ref{xmm_lines} shows the lines as they would be detected with XMM-Newton/RGS1 synthesized from our reference model (Run 4, M3-E3-D6-Z4.5-X4-W1) at four representative epochs, while the lines as they would be detected with XRISM/Resolve at three representative epochs (corresponding to times when XRISM can register a detectable signal) are shown in Fig.~\ref{xrism_lines}.

Figures~\ref{xmm_lines} and \ref{xrism_lines} disentangles the relative contributions to the total X-ray emission from the shocked ejecta (characterized by enhanced metallicity) and the shocked CBM (including disk, EDE, and RG wind), whose properties are shaped by the pre-existing circumbinary structure of the system. By following the temporal evolution of these spectral features, we assess how high-resolution X-ray spectroscopy with XMM-Newton/RGS and XRISM/Resolve can probe the density distribution and geometry of the CBM, while simultaneously tracing the enrichment of the X-ray emitting plasma by freshly shocked ejecta. These synthetic diagnostics may provide a roadmap for interpreting the forthcoming outburst of \tcrb, helping to disentangle the roles of shocked ejecta and shocked CBM in shaping the observable X-ray emission.

The figures show that, in the early phase (a few days after the eruption), the emission in lines below 3\,keV is dominated by shocked ejecta, while higher-energy lines (such as Fe\,XXV at 6.7\,keV) originate primarily from shocked CBM, particularly the dense disk. At later times, shocked ejecta briefly dominates across the energy range, before being overtaken (about one month post-eruption) by emission from shocked CBM, reflecting the growing contribution from the shocked EDE material. The timing and relative contributions of these plasma components vary between models, as illustrated by comparing the spectra synthesized from the reference model (Figs.~\ref{xmm_lines} and \ref{xrism_lines}) with those synthesized from model M3-E3-D8-Z4.5-X8-W1 (Run 3; Figs.~\ref{lines_run3} and \ref{lines_xrism_run3} in Appendix~\ref{app:add_spec}).

The synthetic spectra in Figs.~\ref{xmm_lines} and \ref{xrism_lines} illustrate not only the temporal evolution of line flux and width, but also reveal marked changes in line profile structure. During the earliest epochs (within the first 6 days post-eruption; top two rows of Fig.~\ref{xmm_lines}), soft X-ray lines (particularly O\,VII and O\,VIII) appear broad and asymmetric, with extended blue wings. This asymmetry arises from absorption of redshifted emission by dense, cooler material along the LoS, which preferentially suppresses emission from the receding side of the ejecta. As the remnant evolves (from $\sim 1$ month to 1 year), the lines gradually become narrower and more symmetric. This transition reflects shock deceleration and a decreasing optical depth as the remnant expands, reducing LoS absorption and allowing more balanced emission from both approaching and receding material. These spectral features provide valuable diagnostics of the remnant’s evolving structure, dynamics, and composition.

Importantly, the figures also highlight systematic differences between the contributions of the shocked ejecta and the shocked CBM. The shocked ejecta component tends to produce broader and more structured line profiles, reflecting the higher initial velocities and internal inhomogeneities of the ejected material. In contrast, the emission from the shocked CBM appears in the form of narrower and smoother profiles, as it traces the more uniform and slower-moving plasma shaped by the circumbinary environment. This is particularly evident in the early stage (a few days after the eruption), when most of the emission in lines below 3\,keV originates from the shocked ejecta (upper panels in Fig.~\ref{xmm_lines}). In contrast, lines at higher energies (e.g., Fe\,XXV at 6.7\,keV in Fig.~\ref{xrism_lines}) show a clear dominance of CBM emission, resulting from the substantial amount of shocked material in the accretion disk, as also reflected in the hard X-ray lightcurve (lower left panel in Fig.~\ref{lightcurve}). The transition from ejecta-dominated to CBM-dominated emission over time, combined with the gradual disappearance of asymmetric features, may offer a powerful diagnostic to disentangle the physical properties and geometry of the ejecta and surrounding medium, and to trace how the nova remnant of \tcrb\ evolves in the aftermath of the eruption.

\section{Discussion}
\label{sec:discussion}

\subsection{Constraining the ejected mass and explosion energy}
\label{sec:constr_mass_en}

A key goal of our modeling is to determine whether upcoming X-ray observations of \tcrb\ can constrain the explosion parameters: ejected mass ($M_{\rm ej}$) and explosion energy ($E_{\rm bw}$). Along with the accreted mass needed for thermonuclear runaway (estimated from the WD mass; e.g., \citealt{Drake2016}), these parameters are crucial for understanding the remnant’s evolution and the WD’s long-term fate: whether it gains or loses mass over successive outbursts, potentially reaching the Chandrasekhar limit and exploding as a Type Ia SN or collapsing into a neutron star.

To address this, we examined how variations in ejected mass and the explosion energy affect the predicted X-ray lightcurves, isolating their influence from other environmental factors such as the circumbinary density structure. In this section, we quantify the diagnostic power of X-ray observations by analyzing trends in the synthetic lightcurves across the parameter space explored (see Table~\ref{tab2}).

\begin{figure*}
   \centering
   \includegraphics[width=0.9\textwidth]{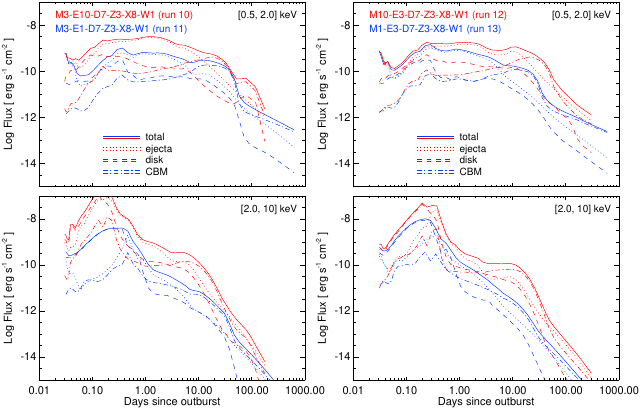}
   \caption{X-ray lightcurves synthesized from the HD models in the $[0.5,2]$~keV (upper panels) and $[2,10]$~keV (lower panels) energy bands. Left panels: comparison between two models with different explosion energies (Runs~10 and~11). Right panels: comparison between two models with different ejecta masses (Runs~12 and~13). The dotted, dashed, and dot-dashed lines show the contributions to the total emission from the shocked ejecta, shocked disk material, and shocked CBM (EDE and RG wind) plasma, respectively.}
   \label{lightcurve_2}%
   \end{figure*}

Figure~\ref{lightcurve_2} presents X-ray lightcurves for a subset of models that differ only in either $M_{\rm ej}$ or $E_{\rm bw}$, while all other parameters remain fixed. In the left panels, we focus on models M3-E10-D7-Z3-X8-W1 (Run 10) and M3-E1-D7-Z3-X8-W1 (Run 11) with fixed $M_{\rm ej} = 3 \times 10^{-7}$~\msun\ and varying explosion energies from $10^{43}$ to $10^{44}$~erg. In the right panels, we compare models M10-E3-D7-Z3-X8-W1 (Run 12) and M1-E3-D7-Z3-X8-W1 (Run 13), which span an ejecta mass range from $10^{-7}$~\msun\ to $10^{-6}$~\msun\, all with $E_{\rm bw} = 3 \times 10^{43}$~erg. 

Our results show that both $M_{\rm ej}$ and $E_{\rm bw}$ exert a strong influence on the X-ray luminosity during the early evolution of the remnant (first $\leq 10$~days), particularly in the hard X-ray band ($[2, 10]$~keV). At early times, a higher explosion energy leads to a faster expansion and stronger shocks, resulting in brighter and harder X-ray emission. Conversely, increasing the ejected mass while keeping the energy fixed reduces the post-shock temperatures (due to energy being distributed over a larger mass), delaying the onset of bright X-ray emission and softening the overall spectrum.

These effects are most pronounced in the hard X-ray band, where peak fluxes vary by up to two orders of magnitude between models with the lowest and highest explosion energy (see lower left panel of Fig.~\ref{lightcurve_2}), and by up to $\approx 1$ order of magnitude between models spanning the lowest to highest ejecta masses (see lower right panel of Fig.~\ref{lightcurve_2}). Furthermore, the lightcurve peak occurs earlier and is sharper in models with higher $E_{\rm bw}$. In the soft band ($[0.5, 2]$~keV), differences are more moderate but still systematic, with peak luminosity varying mainly with $E_{\rm bw}$.

Importantly, these trends persist despite the complexity introduced by the CBM, suggesting that early-time X-ray observations (especially within the first week of the outburst) are particularly well-suited for constraining explosion parameters. In our simulations, models with $M_{\rm ej} = 10^{-6}$~\msun\ produce harder spectra and more luminous emission than those with $M_{\rm ej} = 10^{-7}$~\msun, assuming comparable explosion energies. Similarly, high-energy models (e.g., $E_{\rm bw} = 10^{44}$~erg) display broader, more prolonged high-energy emission due to the contribution of shocked EDE, potentially serving as a distinctive signature of a particularly energetic event.

From a practical standpoint, observations with XRISM/Resolve in the $[2, 10]$~keV range during the first 1–5 days after the eruption will be especially informative, as this is when differences among models are most pronounced and when the contribution from the shocked disk and early shocked ejecta is largest. XMM-Newton, while more sensitive in the soft band, may be more effective at later times, when the shocked ejecta and EDE dominate the emission.

In conclusion, early X-ray lightcurves offer a powerful means to constrain the ejected mass ($M_{\rm ej}$) and explosion energy ($E_{\rm bw}$), though these parameters are often subject to degeneracy: both with each other and with the properties of the accretion disk. This degeneracy can be partially broken by jointly analyzing soft and hard X-ray bands and by tracking the evolution of line diagnostics (see Sect.~\ref{sec:spectra}). In particular, deriving the EM-weighted electron temperature provides an additional constraint: the combined information from plasma flux and temperature helps disentangle the contributions of the disk (primarily its density) from those of the explosion itself. This approach was proved to be especially effective in the case of SN 2014C (\citealt{2024ApJ...977..118O}). A coordinated observing campaign that captures the first hours to days of the \tcrb\ eruption, with high spectral resolution and broad energy coverage, will therefore be essential to fully break these degeneracies and robustly determine the explosion physics.

\subsection{Comparison with RS\,Oph and V745\,Sco}
\label{sec:comp_rsoph_v745}

The X-ray emission from \tcrb, as predicted by our 3D HD models, shows significant similarities and key differences when compared to analogous 3D HD simulations of the well-observed eruptions of RS\,Oph (2006) and V745\,Sco (2014), two prototypical symbiotic recurrent novae.

In RS\,Oph, the hard X-ray flux ($[2, 10]$~keV) peaked within 1 day of the eruption with luminosities of up to $L_{\rm X} \sim 10^{36}$~erg~s$^{-1}$  (\citealt{2006ApJ...652..629B, 2006Natur.442..276S}). The emission declined rapidly, falling by nearly an order of magnitude over the next 10–15 days, with the hard component becoming negligible after $\sim 20$ days. In the soft X-ray band ($[0.3, 2]$~keV), a supersoft source (SSS) phase appeared around day 29, lasting approximately 30–40 days (\citealt{2011ApJ...727..124O}), powered by residual nuclear burning on the WD surface.

V745~Sco showed an even faster evolution: the hard X-ray peak occurred at $\sim 0.5$ days post-eruption, with $L_\mathrm{X} \sim 10^{35.5} - 10^{36}\,\mathrm{erg\,s^{-1}}$ \citep{2015MNRAS.454.3108P, Drake2016}. The emission declined within a few days, becoming undetectable by day 6–8. The SSS phase was either extremely brief or entirely absent, likely due to low accreted mass and strong absorption.

In our \tcrb\ reference model (Run 4, M3-E3-D6-Z4.5-X4-W1), the hard X-ray emission peaks around 5 hours post-outburst, reaching $L_\mathrm{X} \approx 10^{36}\,\mathrm{erg\,s^{-1}}$ (see Fig.~\ref{lightcurve}), driven by the interaction between the ejecta and the dense accretion disk. In models without a disk, the hard X-ray emission is over two orders of magnitude weaker during the first few hours, although a peak still occurs around the same time. This hard component fades rapidly within 1–2 days. The soft X-ray emission evolves more gradually\footnote{Our models do not include the SSS contribution (see Appendix~\ref{app:synthesis}).}. In models with compact EDEs, the soft peak extends to 10 days, whereas in denser or more extended EDE configurations (e.g., Run 3, M3-E3-D8-Z4.5-X8-W1), the peak is delayed to 20–30 days and can persist up to one month after the explosion. As a result, \tcrb\ may show a more extended soft X-ray phase than RS~Oph or V745~Sco. However, it is important to note that these differences primarily reflect our assumed CBM configurations rather than confirmed conditions in \tcrb. The EDE structure in our models is based on several assumptions, and until observational constraints become available, any comparison across systems must be interpreted with caution.

\begin{table}
\centering
\caption{Comparison of X-ray properties for RS~Oph, V745~Sco, and \tcrb.}
\vspace{0.5em}
\label{tab3}
\begin{tabular}{lccc}
\hline\hline
Nova & Peak $L_\mathrm{X}$ & Hard X-ray & Soft X-ray  \\
     & (0.3–10\,keV) & Peak Time & Peak Time  \\
\hline
RS~Oph & $\sim 10^{36}$ erg~s$^{-1}$ & $\sim$1 day & $\sim$30 days \\
V745~Sco & $\sim 10^{36}$ erg~s$^{-1}$ & $\sim$0.5 days & $<$10 days \\
\tcrb\  & $\sim 10^{36}$ erg~s$^{-1}$ & $\sim 0.2$ days & $10-30$ days  \\
\hline
\end{tabular}
\label{tab:comparison}
\end{table}

Table~\ref{tab3} presents a comparison of the X-ray properties of RS~Oph, V745~Sco, and \tcrb, highlighting the critical influence of circumbinary density and geometry on nova outburst observables. RS~Oph and V745~Sco evolve in denser environments, leading to luminous ($L_\mathrm{X} \sim 10^{36}\,\mathrm{erg\,s^{-1}}$) but shorter-lived X-ray emission. In contrast, \tcrb\ reaches a comparable peak luminosity ($L_\mathrm{X} \approx 10^{36}\,\mathrm{erg\,s^{-1}}$), but does so through interaction with a denser accretion disk rather than a uniformly dense ambient medium. Its lower overall circumbinary density and the presence of a more extended and structured EDE contribute to a softer, longer-lasting, and morphologically more complex X-ray signature.

\section{Summary and conclusions}
\label{sec:summary}

In this work, we have estimated unprecedented constraints on the properties of the CBM in the symbiotic recurrent nova \tcrb, and presented the first fully 3D HD simulations aimed at predicting the X-ray signatures of its forthcoming outburst. We have modeled inter-eruption radio observations of \tcrb\ to constrain the density of a spherical wind-like component and a torus-like EDE. We then have modeled the eruption itself, incorporating the complex circumbinary environment (including the RG companion, the accretion disk, a spherical wind, and an EDE). Our HD models explore a broad parameter space spanning explosion energies, ejecta masses, and CBM configurations. We have synthesized the expected X-ray emission as it would be observed by XMM-Newton/RGS and XRISM/Resolve, providing a set of predictions that may be useful for the planning and interpretation of upcoming observations of the nova eruption.

Our principal findings can be summarized as follows:
\begin{itemize}

\item Density of the CBM. As described in Sect.~\ref{sec:cbm}, modelling of \tcrb's inter-eruption radio emission as thermal free-free radiation implies low CBM densities, with a spherical wind component characterized by $\dot{M} = 4 \times 10^{-9}$~M$_{\odot}$~yr$^{-1}$ (for an assumed wind velocity, $v_{\infty} = 10$~km~s$^{-1}$) and a low-density EDE ($n_{\rm ede} \approx 10^6$ cm$^{-3}$), based on the observations presented in \cite{2019ApJ...884....8L}. This is significantly lower density than the CBM in other symbiotic recurrent novae (i.e., RS~Oph, V745~Sco), with likely profound impacts on the multi-wavelength signatures of \tcrb's imminent eruption. The primary way to ``hide" a dense CBM in \tcrb\ would be if it is partially neutral, but optical observations contemporaneous with radio constraints imply the CBM is entirely ionized during the ``super-active" accretion state in question \citep{2016NewA...47....7M}.

\item Blast wave morphology and shaping by the CBM. As detailed in Sect.~\ref{sec:hydro}, the blast wave evolves into a distinctly bipolar morphology, collimated by the accretion disk and the dense EDE, and further sculpted by the RG companion. Figures~\ref{evol} and \ref{models} illustrate that the blast propagates more rapidly along the polar directions, while decelerating in the denser equatorial plane. The RG companion induces a prominent bow shock on the leading side and a hot wake downstream, generating persistent asymmetries lasting several weeks. Variations in the density and scale height of the EDE critically affect both the degree of collimation and the overall spatial extent of the remnant.

\item X-ray lightcurves and temporal evolution of EM distributions. The synthetic X-ray lightcurves (Fig.~\ref{lightcurve_all}) reveal three distinct evolutionary phases: (i) an initial phase lasting several hours, dominated by hard X-ray emission from shocked disk material; (ii) an intermediate phase spanning up to a few weeks or about one month, contingent on the EDE properties, dominated by shocked ejecta; and (iii) a late phase starting thereafter, where the shocked CBM becomes the principal source of X-ray emission. These lightcurves provide a powerful diagnostic tool for disentangling the contributions of different plasma components and constraining the physical structure of the circumbinary environment, including the presence and density of an accretion disk or EDE. EM distributions (Fig.~\ref{em_dist}) further support this interpretation, confirming the presence of multi-temperature plasma and revealing significant departures from CIE, particularly at late times.

\item X-ray emission morphology. As shown in Fig.~\ref{emission_maps}, the remnant’s X-ray morphology undergoes pronounced changes over time. Initially, emission is concentrated near the interaction region between the blast wave and the disk. Within days, a ring-like emission structure emerges due to the dense EDE. At later epochs, the remnant adopts a bipolar morphology consistent with easier blast expansion along the lower-density polar directions.

\item Spectral diagnostics and line profile evolution. We synthesized XMM-Newton/RGS and XRISM/Resolve spectra at key epochs (Figs.~\ref{xmm_lines}, \ref{xrism_lines}). The total X-ray flux gradually declines over time, especially at energies $>1$\,keV, resulting in a progressively softer spectrum (see lightcurves in Sect.~\ref{sec:lightcurves}). In the reference model, early spectra (up to $\sim 1$ month, depending on EDE properties) are dominated by shocked ejecta, producing broad, structured, and asymmetric lines with extended blue wings due to absorption of redshifted emission along the LoS. At later times, shocked CBM dominates, yielding narrower, smoother profiles. As the remnant expands and the absorbing material becomes more diffuse, line asymmetries diminish. Line profiles from shocked ejecta systematically differ from those of shocked CBM, reflecting higher velocities and more complex structure in the former.

\end{itemize}

In light of our results, we conclude that the temporal evolution of line fluxes, widths, and asymmetries offers powerful diagnostics to probe the density distribution and geometry of the CBM, as well as the kinematics and composition of the ejecta during the next outburst of \tcrb. High-resolution and time-resolved X-ray observations will be key to capturing these signatures and characterizing the structure and evolution of the nova remnant. Instruments like XRISM/Resolve are well suited to detect high-energy line features sensitive to shock dynamics and plasma ionization states, while XMM-Newton/RGS provides excellent sensitivity at lower energies, ideal for identifying emission from shocked ejecta. Complementarily, instruments such as XMM-Newton/EPIC and XRISM/Xtend offer broad energy coverage and high time resolution, crucial for tracking the evolving lightcurves. The combined capabilities of high-resolution microcalorimeters and CCDs will therefore be essential for a comprehensive, multi-faceted characterization of \tcrb's forthcoming eruption, directly linking theoretical predictions to observational data.

A comparison between the past outbursts of RS~Oph and V745~Sco and our model predictions for the forthcoming eruption of \tcrb\ highlights the decisive role of circumbinary density and geometry in shaping the observable signatures of nova outbursts. RS~Oph and V745~Sco, embedded in denser environments, produce luminous ($L_\mathrm{X} \sim 10^{36}$ erg s$^{-1}$) but short-lived X-ray emission. In contrast, our simulations suggest that \tcrb\ may reach a comparable peak luminosity, primarily through interaction with a dense accretion disk rather than a uniformly dense ambient medium. Assuming a lower overall circumbinary density and the presence of an extended, structured EDE, the resulting shock evolution in \tcrb\ is expected to give rise to a softer, longer-lasting, and morphologically more complex X-ray signal. While these predictions rely on a set of assumptions regarding the CBM configuration and system parameters, they illustrate how variations in circumbinary geometry can fundamentally alter nova remnant evolution and provide valuable diagnostics for interpreting outbursts in symbiotic systems.

Building on these insights, our 3D simulations provide a predictive framework for the forthcoming eruption of \tcrb. Once the nova event occurs, comparison between observational data and our synthetic predictions will allow for model calibration and validation. As demonstrated in past work on SNe (e.g., SN 1987A, \citealt{2025A&A...699A.305O} and SN 2014c, \citealt{2024ApJ...977..118O}), such an approach enables robust constraints on both explosion parameters and the surrounding medium. Upon validation, the models developed here may serve as a generalizable tool for interpreting nova–CBM interactions, with implications for recurrent novae, Type Ia SN progenitors, and the long-term morphological evolution of nova remnants. Future coordinated X-ray campaigns, coupled with multi-wavelength follow-up, will be essential to maximize the scientific return and refine our understanding of nova evolution in complex binaries.

Finally, novae are now routinely detected in GeV $\gamma$-rays \citep{Chomiuk+21}, and RS\,Oph was recently identified as the first nova to exhibit TeV $\gamma$-ray emission \citep{2022NatAs...6..689A, 2022Sci...376...77H, 2025A&A...695A.152A}. Given the similarities between \tcrb\ and RS\,Oph (particularly in terms of their binary configurations and dense CBM) the potential for high-energy emission from \tcrb\ is compelling (e.g., \citealt{2025arXiv250707096T}). The CBM structures modeled in this work provide a foundation for a companion paper (Petruk et al., in preparation) aimed at predicting the associated $\gamma$-ray and neutrino signals in the event of a future outburst.

\begin{acknowledgements}
We thank an anonymous referee for the useful suggestions that allowed us to improve the manuscript. The \PLUTO\ code is developed at the Turin Astronomical Observatory (Italy) in collaboration with the Department of General Physics of  Turin University (Italy) and the SCAI Department of CINECA (Italy). We acknowledge that part of the results of this research have been achieved using the HPC system MEUSA at the SCAN (Sistema di Calcolo per l'Astrofisica Numerica) facility for HPC at INAF-Osservatorio Astronomico di Palermo. S.O., M.M., F.B., and O.P. acknowledge financial contribution from the PRIN 2022 (20224MNC5A) - ``Life, death and after-death of massive stars'' funded by European Union – Next Generation EU and from the INAF Theory Grant ``Supernova remnants as probes for the structure and mass-loss history of the progenitor systems''. 
OP acknowledges the support from INAF 2023 RS4 mini-grant.
M.M. acknowledges support by the  Fondazione ICSC, Spoke 3 Astrophysics and Cosmos Observations. National Recovery and Resilience Plan (Piano Nazionale di Ripresa e Resilienza, PNRR) Project ID CN\_00000013 "Italian Research Center on  High-Performance Computing, Big Data and Quantum Computing" funded by MUR Missione 4 Componente 2 Investimento 1.4: Potenziamento strutture di ricerca e creazione di "campioni nazionali di R\&S (M4C2-19 )" - Next Generation EU (NGEU). L.C. is grateful for support from NASA grants 80NSSC23K0497, 80NSSC23K1247, and 80NSSC25K7334, along with National Science Foundation grant AST-2107070. 
The National Radio Astronomy Observatory is a facility of the U.S. National Science Foundation operated under cooperative agreement by Associated Universities, Inc.
\end{acknowledgements}

\bibliographystyle{aa}
\bibliography{references}

\clearpage
\begin{appendix}

\section{Simulation setup, numerical methods, and grid strategy}
\label{app:setup}

The numerical setup is based on those previously developed to model the HD evolution of SN 1987A \citep{2015ApJ...810..168O, 2019A&A...622A..73O, 2020A&A...636A..22O} and SN~2014c (\citealt{2024ApJ...977..118O}), systems characterized by a complex, highly structured circumstellar environment and pronounced asymmetries. We solve the time-dependent equations of mass, momentum, and energy conservation using the \PLUTO\ code \citep{2007ApJS..170..228M, 2012ApJS..198....7M}, a modular, Godunov-type framework for computational astrophysics. The HD equations are solved using the Roe approximate Riemann solver, combined with a third-order Runge–Kutta time integration scheme. Spatial reconstruction is performed with a monotonized central (MC) limiter applied to the primitive variables, the least diffusive option available in \PLUTO.

To account for key physical processes relevant to nova blast wave evolution, we have extended the code with additional computational modules developed for the description of SN blast waves expanding through an inhomogeneous ambient environment. These include electron-ion non-equilibration, implemented through an instantaneous electron heating prescription at shock fronts up to $kT\sim 0.3$~keV via lower hybrid wave dissipation \citep{2007ApJ...654L..69G}, followed by Coulomb-mediated equilibration of electron and ion temperatures in the post-shock plasma \citep[see][]{2015ApJ...810..168O}. Moreover, we track deviations from ionization equilibrium for the most abundant species by computing the maximum ionization age ($n_{\rm e}t$, where $n_{\rm e}$ is the electron number density and $t$ is the time since the plasma was shocked) within each computational cell \citep{2015ApJ...810..168O}. These extensions allow for a physically accurate representation of the evolving thermal and ionization structure of the blast, critical for interpreting multi-wavelength observational diagnostics.

Following the methodology adopted in SN–SNR simulations (e.g., \citealt{2020A&A...636A..22O, 2021A&A...645A..66O}), we employed a remapping technique to track the evolution of the blast wave from a few minutes after the outburst through to the full expansion phase, up to one year post-explosion. The initial computational domain is defined on a Cartesian grid with $(512)^3$ cells, covering a volume of $1.2 \times 10^{12}$~cm per side and achieving a spatial resolution of $4.7 \times 10^{9}$~cm. In a previous study simulating the interaction of SN 2014c with its inhomogeneous circumstellar medium \citep{2024ApJ...977..118O}, we demonstrated that this spatial resolution is sufficient to resolve the complex interaction between the blast wave and the dense structures surrounding the progenitor star. 

As the blast wave propagates outward, the domain is iteratively expanded, with all physical quantities remapped onto the enlarged grid at each step (see \citealt{2020A&A...636A..22O} for details of the remapping procedure). Over the course of one simulated year, approximately 50 remappings were performed, ultimately resulting in a computational domain spanning $9 \times 10^{15}$~cm (corresponding to $\sim 600$~au), with a final spatial resolution of $\approx 3.5 \times 10^{13}$~cm. This approach offers a powerful balance between accuracy and efficiency. By dynamically expanding the computational domain and remapping physical quantities at each stage, we maintain high spatial resolution where and when it is needed (capturing the early-time dynamics near the WD) while avoiding unnecessary computational costs as the blast evolves. This technique allows us to follow the system's development over several orders of magnitude in scale, from the initial compact explosion to the large-scale remnant, ensuring physical fidelity throughout and enabling direct comparison with late-time observations. Boundary conditions were fixed at the pre-explosion CBM values throughout the simulations.

\section{Synthesis of X-ray emission}
\label{app:synthesis}

Based on the results of our 3D HD simulations, we synthesized the X-ray emission arising from the interaction of the blast wave with the surrounding medium, following the methodology established in previous works (e.g., \citealt{2015ApJ...810..168O, 2019NatAs...3..236M, 2024ApJ...977..118O, 2025A&A...699A.305O}). The goal is to predict the X-ray observables of the anticipated eruption of \tcrb\ and to provide a framework for interpreting forthcoming high-resolution, multi-wavelength observations.

In the original simulation setup, the binary system’s equatorial plane lies in the $[x, y]$ plane. To match the observed inclination of the \tcrb\ system \citep{2025ApJ...983...76H}, we applied a $55^\circ$ rotation about the $x$-axis (see Table~\ref{tab1}). An additional $20^\circ$ rotation about the $z$-axis was applied beforehand to represent a generic orbital phase at the time of outburst (this value may vary depending on the actual eruption time). These transformations align the simulated geometry with the observer's LoS, assumed to be along the negative $y$-axis, thereby ensuring a consistent comparison with observational data.

We evaluated several key plasma properties for each computational cell to characterize the X-ray emission throughout the simulation domain. The ionization age was calculated as $\tau_j = n_{\rm e,j} \Delta t_{\rm j}$, where $n_{\rm e,j}$ is the electron number density and $\Delta t_{\rm j}$ the time since shock passage in the $j$-th cell. The EM was computed as ${\rm EM}j = n_{\rm e,j} n_{\rm Z,j} V_j$, with $n_{\rm Z,j}$ the ion number density and $V_j$ the volume for the $j$-th cell. The electron temperature $T_{\rm e,j}$ was initialized at a post-shock value of $kT = 0.3$\,keV, consistent with rapid heating via lower hybrid waves \citep{2007ApJ...654L..69G, 2015ApJ...810..168O}. Its subsequent evolution was calculated by modeling Coulomb energy exchange between ions and electrons, based on local plasma conditions and the time elapsed since shock heating.

The chemical abundances of the CBM around \tcrb\ are not well constrained. Thus, following \citet{2017MNRAS.464.5003O}, we adopted the solar abundances of \citet[][hereafter AG]{1989GeCoA..53..197A} for the RG wind, the accretion disk, and the EDE. More recent determinations of solar abundances are available (e.g., \citealt{2003ApJ...591.1220L, 2025SSRv..221...23L}), but given the uncertainty in the CBM composition, we chose to retain the AG abundances used in previous studies (e.g., \citealt{2017MNRAS.464.5003O}). In fact, this approach ensures consistency with these studies and facilitates a direct comparison between the new simulations and earlier results (as, for instance, in Sect.~\ref{sec:comp_rsoph_v745}). For the nova ejecta, we assumed AG abundances enhanced by a factor of 10, motivated by the detection of metal-rich material in high-resolution X-ray spectra of RS\,Oph and V407\,Cyg (e.g., \citealt{2009ApJ...691..418D, 2011A&A...527A..98S}). X-ray emission in the $[0.5, 10]$ keV band was synthesized using the non-equilibrium ionization (NEI) model VPSHOCK within the XSPEC package, with atomic data from ATOMDB \citep{Smith2001ApJ,Foster2020}. The resulting spectra include the final ionization fractions as predicted by the NEI model, based on the computed values of $\tau_j$, EM$_j$, and $T_{\rm e,j}$. A source distance of 890 pc was assumed, consistent with the GAIA-inferred value of $887^{+22}_{-23}$\,pc \citep{2021AJ....161..147B}.

To model realistic spectra, we included Doppler shifts of emission lines due to bulk plasma velocities along the LoS, following the method described in \citet{2009A&A...493.1049O}. Absorption by the interstellar medium (ISM), the local CBM, and the ejecta was also accounted for. Local photoelectric absorption was computed self-consistently from the simulation’s density and composition distributions, using wavelength-dependent cross-sections from \citet{1992ApJ...400..699B}. Interstellar absorption was modeled with a neutral hydrogen column density of $N_{\rm H} = 5 \times 10^{20}$ cm$^{-2}$, consistent with the 2D HI4PI all-sky H~I survey \citep{2016A&A...594A.116H}.

The synthetic spectra were convolved with the instrumental responses of XMM-Newton/RGS and XRISM/Resolve, assuming an exposure time ranging between $t_{\rm exp} = 10$ ks and $t_{\rm exp} = 500$ ks, possible for observations with these instruments. This integration time is important to capture time-averaged spectral broadening due to the remnant's dynamical evolution during the observation. In XRISM/Resolve spectra, we note that emission below 1.8\,keV is significantly attenuated due to the gate valve anomaly, which failed to open as originally planned. This technical issue greatly reduces the instrument’s sensitivity at softer energies and decreases the effective area at harder energies, thereby constraining the available diagnostics.

We did not simulate Chandra spectra (HETG-based) primarily because the ACIS detector has suffered severe molecular contamination on its optical blocking filter, resulting in substantial loss of effective area in the soft X‑ray band where \tcrb’s emission is strongest\footnote{https://cxc.cfa.harvard.edu/ciao/why/acisqecontamN0015.html} (below $\sim 1$~keV). While it is technically possible to operate the HETG with the HRC detector to mitigate ACIS contamination, that configuration is generally not recommended: the HRC lacks energy resolution for order sorting, the background is higher, and the overall effective area above $\sim 1$~keV is limited, making grating spectra much less efficient or reliable in the soft band for faint, soft sources like \tcrb. 

This modeling framework captures the full 3D evolution of the nova remnant and allows us to predict its X-ray morphology, spectral features, and variability. By systematically exploring a range of explosion energies, ejecta masses, and CBM configurations, we aim to constrain the physical conditions of the \tcrb\ system and provide observational templates to guide and interpret upcoming multi-wavelength campaigns.

It is important to note that our emission synthesis specifically focuses on the thermal X-ray emission produced by shocks propagating through the ejecta and the surrounding CBM. As such, it does not currently include the contribution from the super-soft source (SSS) phase, which originates from residual nuclear burning on the WD surface following the outburst. This choice is justified because the SSS emission, which typically peaks at very soft X-ray energies (below $\sim 0.5$~keV), primarily traces the atmospheric emission of the WD itself and is not directly related to the shock-heated plasma shaping the nova remnant. While excluding the SSS component may lead to an underestimation of the total emission at the lowest energies (particularly in the early weeks after eruption) it does not significantly affect the characterization of the shock-driven emission, which encodes key information about the nova’s interaction with its environment. If needed, the SSS component can be included in the synthetic spectra to provide a more comprehensive view of the nova’s X-ray evolution, especially in the very soft band, and to help disentangle the contributions from nuclear burning and shock-heated material.

\section{Emission measure distribution versus temperature and ionization time}
\label{sec:em_dist}

   \begin{figure*}[!h]
   \centering
   \includegraphics[width=\textwidth]{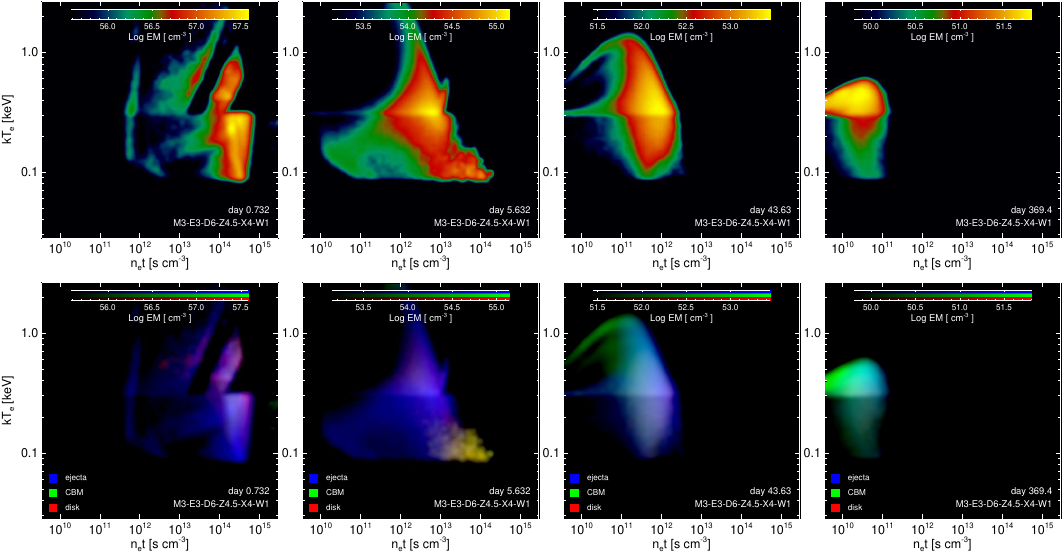}
   \caption{Upper panels: EM distributions as a function of electron temperature ($kT_{\rm e}$) and ionization parameter ($\tau = n_{\rm e}t$) at the indicated times for model M3-E3-D6-Z4.5-X4-W1 (Run 4). Lower panels: Corresponding three-color composite maps showing the distribution of EM. Colors indicate the contribution from different shocked plasma components: ejecta (blue), CBM (without the disk; green), and accretion disk (red).}
   \label{em_dist}%
   \end{figure*}


The HD models presented in Sect.~\ref{sec:hydro} offer a detailed and physically consistent framework for investigating the shock-heated plasma generated by the forthcoming nova eruption of \tcrb. In particular, they allow us to track the evolution of the emission measure (EM) distribution as a function of electron temperature ($kT_{\rm e}$) and ionization parameter ($\tau$), providing key diagnostics of the thermal and ionization states of the shocked material over time. Figure~\ref{em_dist} shows the temporal evolution of the EM distribution for our reference model, M3-E3-D6-Z4.5-X4-W1 (Run 4). The other models exhibit qualitatively similar trends. In the following, when referring to the contribution from the CBM, we implicitly exclude the contribution from the accretion disk which is analyzed separately.

As anticipated in Sect.~\ref{sec:hydro}, the EM distributions reveal a complex, multi-temperature structure that evolves significantly throughout the simulation and spans a temperature range from $\sim 0.1$~keV to $\sim 1$~keV and a ionization parameter spanning from $10^{10}$~s~cm$^{-3}$ to $10^{15}$~s~cm$^{-3}$. These results indicate that the X-ray–emitting plasma can substantially deviate from collisional ionization equilibrium (CIE), especially during the later stages of the remnant’s evolution, beyond one month after the outburst. Both shocked ejecta and shocked circumbinary material, including contributions from the accretion disk and the CBM (EDE and RG wind), are responsible for the X-ray emission, as shown in the bottom panels of Fig.~\ref{em_dist}. Although these components follow distinct evolutionary tracks, their EM distributions show partial overlap throughout the simulation.

In the early phase, shortly after the blast wave encounters the dense disk and surrounding CBM, the EM distribution exhibits a prominent peak at $kT_{\rm e} \approx 0.2-0.3$~keV with $\tau > 10^{13}$~s~cm$^{-3}$, suggesting that the plasma is in CIE due to the high densities encountered by the shock. During this phase, the emission is primarily dominated by the shocked ejecta, with a secondary contribution from shocked material in the disk, particularly evident at higher temperatures (see reddish area in the lower left panel of the figure). As the remnant expands, the shocked plasma begins to interact with lower-density regions, and the EM distribution gradually shifts toward lower values of the ionization parameter, indicating an increasing deviation from CIE. Around one month after the outburst (right-hand panels in Fig.~\ref{em_dist}), the ionization parameter drops below $\tau \sim 10^{12}$~s~cm$^{-3}$, particularly for newly shocked, lower-density material in the CBM and outer ejecta. We observe only the recently shocked material because the plasma at $\sim 0.2$~keV that was previously in CIE has cooled within about a month and no longer emits significantly in X-rays. During this phase, the contribution from the shocked CBM becomes increasingly significant, particularly at higher temperatures, as evident from the growing green regions in the lower panels of Fig.~\ref{em_dist}. This shift reflects the evolving dominance of different plasma components as the blast wave moves through the structured circumbinary environment.

In the late stages of the remnant’s evolution, approaching one year post-eruption, the plasma cools progressively. By the end of the simulation, the shocked material has temperatures below 0.5~keV, with a further decrease in the overall EM, marking the transition to a more quiescent phase. These results highlight the importance of time-dependent, NEI modeling for accurately capturing the X-ray signatures of nova remnants like \tcrb.

In RS~Oph, EM values for the shocked plasma were estimated to be in the range EM$\sim 10^{56} - 10^{57}\,\mathrm{cm^{-3}}$ during the first week \citep{2009ApJ...691..418D, 2009A&A...493.1049O}, declining by roughly an order of magnitude by day 30. Our simulations for \tcrb\ predict EM values of comparable magnitude. In our reference case, we found that: at 0.3 days, the total EM in the shocked ejecta and disk is $\sim 10^{57}\,\mathrm{cm^{-3}}$ (see Fig.~\ref{em_dist}); at 5 days, the EM decreases, reaching $\sim 10^{55}\,\mathrm{cm^{-3}}$; by day 30, the EM continuous to decline gradually. In dense EDE models (e.g., Run 3 in Table~\ref{tab2}), EM values remain above $10^{56}\,\mathrm{cm^{-3}}$ for more than a month. V745~Sco produced higher early EMs ($\sim 10^{57}\,\mathrm{cm^{-3}}$ within the first day; \citealt{Drake2016}), but declined rapidly, becoming undetectable in X-rays within 10–14 days.

\section{XMM-Newton/RGS1 and XRISM/Resolve spectra of model M3-E3-D8-Z4.5-X8-W1 (Run 3)}
\label{app:add_spec}

\begin{figure*}
   \centering
   \includegraphics[width=0.7\textwidth]{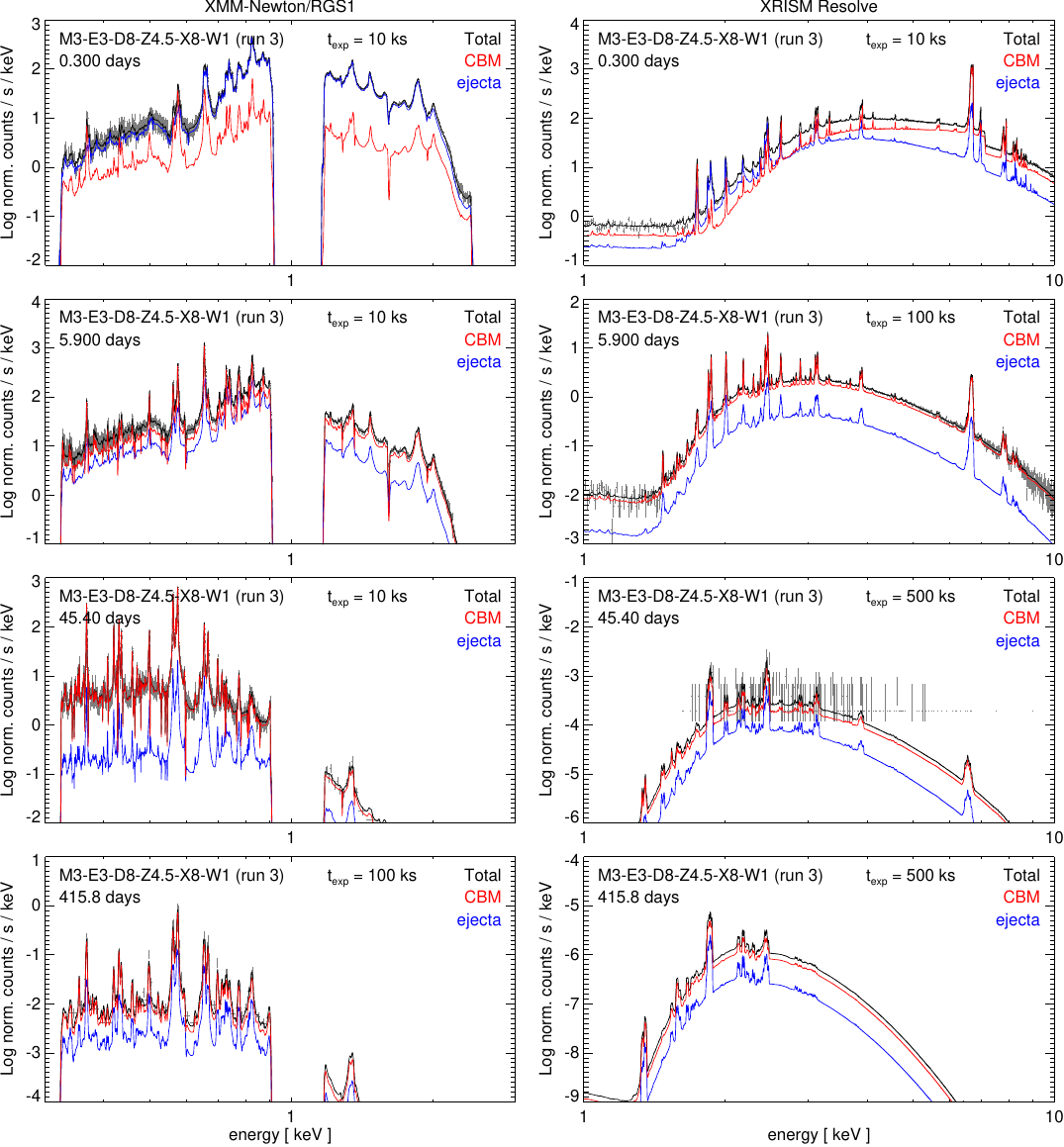}
   \caption{Same as in Fig.~\ref{tot_spectra}, but showing results for model M3-E3-D8-Z4.5-X8-W1 (corresponding to Run~3 in Table~\ref{tab2}).}
   \label{tot_spectra_run3}%
   \end{figure*}

\begin{figure*}
   \centering
   \includegraphics[width=0.95\textwidth]{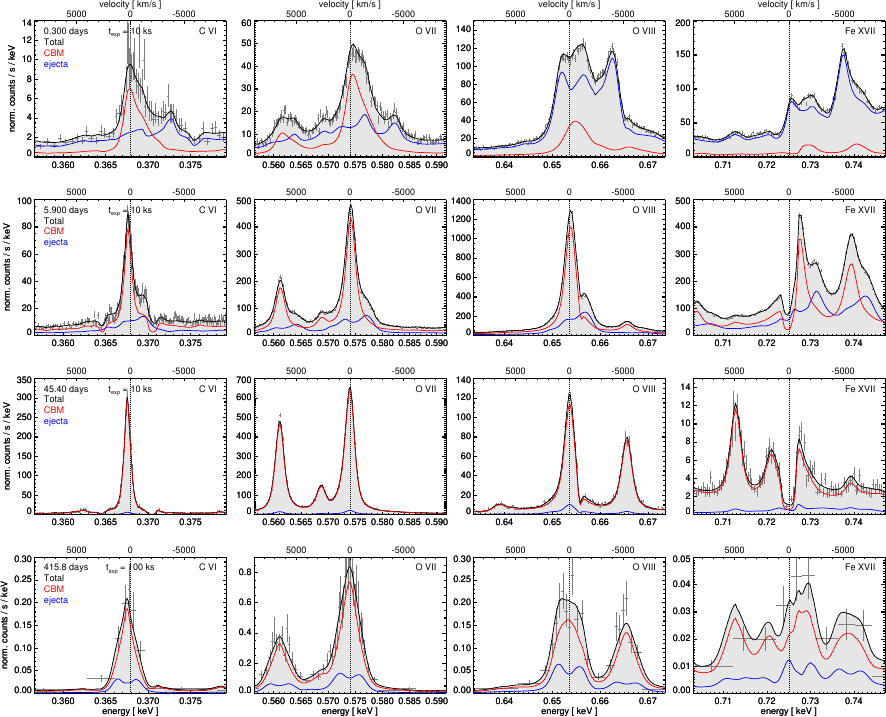}
   \caption{Same as in Fig.~\ref{xmm_lines}, but showing results for model M3-E3-D8-Z4.5-X8-W1 (corresponding to Run~3 in Table~\ref{tab2}).}
   \label{lines_run3}%
   \end{figure*}

\begin{figure*}
   \centering
   \includegraphics[width=0.95\textwidth]{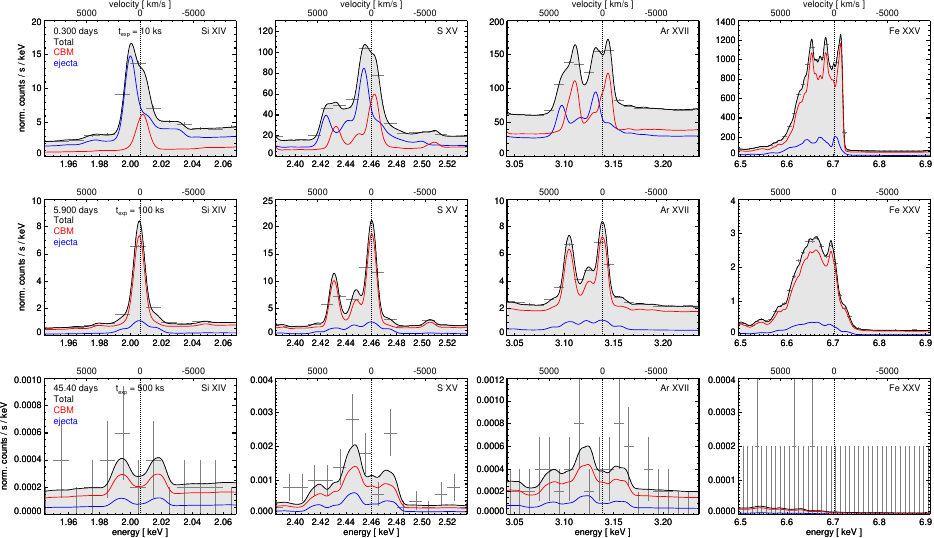}
   \caption{Same as in Fig.~\ref{xrism_lines}, but showing results for model M3-E3-D8-Z4.5-X8-W1 (corresponding to Run~3 in Table~\ref{tab2}).}
   \label{lines_xrism_run3}%
   \end{figure*}

Figure\,\ref{tot_spectra_run3} presents XMM-Newton/RGS1 and XRISM/Resolve spectra derived from model M3-E3-D8-Z4.5-X8-W1 (corresponding to Run\,3 in Table~\ref{tab2}). This model is characterized by a very dense EDE with a peak hydrogen number density of $n_{\rm ede} = 10^8$\,cm$^{-3}$, extending in the equatorial plane with scale lengths $h_{\rm x} = h_{\rm y} = 8 \times 10^{14}$\,cm. The primary difference from the reference model (Run~4; M3-E3-D6-Z4.5-X4-W1) is the substantially larger amount of EDE material that becomes shocked over the time interval considered. According to the simulation, an exposure time of 10~ks is sufficient to obtain XMM-Newton/RGS1 spectra with excellent signal-to-noise ratio during the first month of evolution, while exposure times on the order of 100~ks yield high-quality spectra at later epochs, up to an age of approximately 400~days. For XRISM/Resolve, spectra with good signal-to-noise can be obtained within the first few tens of days with exposure times below 100\,ks. Beyond one month, the signal drops rapidly, and even with a 500\,ks exposure, no significant signal is detected.

Figures~\ref{lines_run3} and \ref{lines_xrism_run3} show synthetic XMM-Newton/RGS1 and XRISM/Resolve line profiles, respectively, for selected lines. As anticipated by the X-ray lightcurves (left panels of Fig.~\ref{lightcurve}), the denser and more extended EDE in this model leads to emission that, in many lines, is dominated by shocked CBM even at the earliest stages, just a few hours after the eruption. Specifically, the shocked CBM becomes the primary source of emission as early as 7~hours after the eruption in the C~VI and O~VII lines (top row of Fig.~\ref{lines_run3}), and likewise in Ar~XVII and Fe~XXV (top row of Fig.~\ref{lines_xrism_run3}). At the same time, the shocked ejecta remain the dominant contributors to the O~VIII, Fe~XVII, Si~XIV, and S~XV lines. A few days later (second row in the figures), the shocked CBM overtakes as the dominant source for all lines, and this trend continues at later times up to the end of the simulation (day~416; fourth row in Fig.~\ref{lines_run3}). This evolution clearly illustrates how the physical properties of the CBM (particularly the density and spatial extent of the EDE) can profoundly affect the X-ray emission, shaping both the relative contributions from shocked ejecta and CBM and the temporal evolution of individual lines.

\end{appendix}

\end{document}